\documentclass[amsmath,amssymb,aps,english,superscriptaddress,nofootinbib]{revtex4}
\usepackage{graphicx,amssymb,amsmath}
\usepackage{babel}
\usepackage{color}
\definecolor{orange}{rgb}{1,0.5,0}
\definecolor{darkorange}{rgb}{0.75,0.25,0}
\definecolor{darkgreen}{rgb}{0,0.4,0}
\definecolor{lightgreen}{rgb}{0,0.7,0}
\definecolor{turquoise}{rgb}{0,0.5,0.5}
\definecolor{lightblue}{rgb}{0.1,0.1,0.8}
\definecolor{darkblue}{rgb}{0.,0.,0.5}
\definecolor{darkviolet}{rgb}{0.5,0.,0.5}
\definecolor{violet}{rgb}{0.7,0.,0.7}
\definecolor{darkred}{rgb}{0.80,0.,0.}

\newcommand{\cl}[1]{{\color{red} #1}}
\newcommand{\cld}[1]{{\color{darkviolet} #1}}

\newcommand{\clt}[1]{{\color{darkgreen} #1}}

\newcommand{\clq}[1]{{\color{darkorange} #1}}
\newcommand{\clqq}[1]{{\color{orange} #1}}
\newcommand{\clc}[1]{{\color{lightgreen} #1}}
\newcommand{\cls}[1]{{\color{darkblue} #1}}
\newcommand{\clss}[1]{{\color{darkred} #1}}
\newcommand{\clh}[1]{{\color{turquoise} #1}}



\usepackage{natbib}

\usepackage{graphicx}
\usepackage{dcolumn}
\begin{document}


\title{Multi-scale analysis of the European airspace using network community detection}



\author{G\'erald Gurtner}
\affiliation{Scuola Normale Superiore, Piazza dei Cavalieri 7, Pisa, 56126, Italy}

\author{Stefania Vitali}
\affiliation{Dipartimento di Fisica e Chimica, Universit\`a degli Studi di Palermo, Viale delle Science Ed. 18, Palermo, 90128, Italy}

\author{Marco Cipolla}
\affiliation{Dipartimento di Fisica e Chimica, Universit\`a degli Studi di Palermo, Viale delle Science Ed. 18, Palermo, 90128, Italy}

\author{Fabrizio Lillo}
\affiliation{Scuola Normale Superiore, Piazza dei Cavalieri 7, Pisa, 56126, Italy}
\affiliation{Dipartimento di Fisica e Chimica, Universit\`a degli Studi di Palermo, Viale delle Science Ed. 18, Palermo, 90128, Italy}
\affiliation{Santa Fe Institute, 1399 Hyde Park Road, Santa Fe, NM 87501}

\author{Rosario Nunzio Mantegna}
\affiliation{Dipartimento di Fisica e Chimica, Universit\`a degli Studi di Palermo, Viale delle Science Ed. 18, Palermo, 90128, Italy}
\affiliation{Central European University, Center for Network Science and Department of Economics, Nador u. 9, 1051 Budapest, Hungary}
\author{Salvatore Miccich\`e}
\affiliation{Dipartimento di Fisica e Chimica, Universit\`a degli Studi di Palermo, Viale delle Science Ed. 18, Palermo, 90128, Italy}

\author{Simone Pozzi}
\affiliation{Deep Blue s.r.l., Piazza Buenos Aires 20, 00100 Roma, Italy.}

\begin{abstract}
We show that the European airspace can be represented as a multi-scale traffic network whose nodes are airports, sectors, or navigation points and links are defined and weighted according to the traffic of flights between the nodes. By using a unique database of the air traffic in the European airspace, we investigate the architecture of these networks with a special emphasis on their community structure. We propose that unsupervised network community detection algorithms can be used to monitor the current use of the airspaces and improve it by guiding the design of new ones. 
Specifically, we compare the performance of three community detection algorithms, also by using a null model which takes into account the spatial distance between nodes, and we discuss their ability to find communities that could be used to define new control units of the airspace. 
\end{abstract}

\date{\today}

\maketitle

\section{Introduction}

The application of network theory to air traffic is not new and many papers have been already published on the subject (for a recent review, see \cite{ZaninLillo13}). These studies have focused mainly on the topological characterization of the airport network \cite{LiPing03,Barrat04,Li04,Guimera05,Barrat05, Guida07,Bagler08,Xu08,Quartieri08,Zanin08,Xu11,Wang11,Lillo11,Vinko12}. In this network, airports are nodes and a link exists if two airports are connected by a direct flight. Often the number of flights between two airports in a given time window is used to weight the links, making the graph an instance of a traffic network. The interest in airport networks comes from the need of modelling traffic flow, mobility of passengers, and spreading of infectious diseases \cite{Colizza06}.

An important characteristic of a complex network is its organization in communities (clusters) \cite{Fortunato10}. Communities are generically defined as sets of nodes that are more connected among themselves than with the rest of the network. Communities are, therefore, an important element to understand and model the architecture of a network. The purpose of this paper is the identification of communities in the different networks that can be defined in the air traffic system.

Airspace is in fact a complex system which is partitioned for a series of reasons, mainly related to air traffic control. As we explain below, the European airspace is partitioned in a hierarchical way. At the highest level, the space is partitioned into multinational areas, termed Functional AirBlocks (FAB). The FABs are not yet fully implemented, but are planned in the near future as a mean to increase the capacity in terms of traffic. Then each country has its own National Airspace (NA), which is typically partitioned into several Air Control Centers (ACC). Each of these is itself partitioned into sectors, which are the smallest unit of control, being under the direct supervision of air traffic controllers. Finally, inside the sectors we find the navigation points constituting the grid where the flights move. In fact, nowadays flight plans are defined as a set of consecutive fixed points that the aircraft is supposed to pass at predefined times. On the smallest scale, therefore, a flight plan is a path on a grid whose nodes are the navigation points. The choice of the boundaries of these multiple partitions is decided in a strongly supervised and not fully quantitative way, taking into account political and strategical reasons and also traffic considerations. 

 To the best of our knowledge this is the first attempt to apply community detection to networks of the air traffic systems. In fact, we do not consider only airport networks but we consider three different types of network of the airspace creating a multi-scale structure. Beside the airport network, we will consider the sector network and the navigation point network. The former is a network where nodes are sectors  -- the smallest units of control -- and links indicate the presence of traffic of flights going from one sector to another. The latter takes the navigation points as nodes, the links being built in the same way.  Making use of a unique and detailed database of the European air traffic, we investigate the topological and community properties of the sector network and of the navigation point network. To the best of our knowledge, the sector network has not been investigated before, while the topological properties of the navigation point network has been investigated only in Ref. \cite{Cai12} in the case of the Chinese airspace.  

As detailed below, we believe that community detection in air traffic networks is important for two reasons. First, it obviously improves the characterization of the networks with respect to analysis based on the measurement of the standard metrics of network topology (degree distribution, betweenness centrality, small world effect, etc.) already considered in the literature, at least for the airport network. Secondly, and more importantly, we believe that community detection could be helpful to guide the design of new airspaces in order to have a better control of the air traffic. In particular, the Single European Sky ATM Research (SESAR) programme -- which aims at a complete reorganization of the Air Traffic Management (ATM) in the next twenty years in answer to the increase of traffic in Europe -- could benefit from this method.

In this paper we show how the community detection in networks provides information on the appropriateness of the airspace design at the various scales considered, based on the sole knowledge of the actual air traffic data. In this respect, 
the added value of this paper is twofold. On one side we show how methods devised for identifying communities in networks could be used to help designing the structure of airspace in a bottom-up way, i.e. starting from the observed behavior of the system. On the other side, this analysis could be seen as an ``horse race" among different community detection methods in order to find which works best when the investigated network describes a traffic flow (not necessarily aircraft, but also cars, data, etc.). These two aspects can not be kept fully distinct. Indeed, it is hard to distinguish whether the actual traffic flow is a consequence of the airspace partition or viceversa. As we mentioned above the airspace partition is due to different reasons sometimes unrelated with the effective air traffic needs. 
  
The paper is organized as follows. In Section \ref{sec:description} we present the current structure of airspace, the considered networks, and their ``natural" community structure, based on the existing partitions (see also the Appendix). Section \ref{sec:data} describes our unique and complex database and in Section \ref{subsec:methods} we present the algorithms for community detection, for comparing partitions, and to characterize the identified communities. Section \ref{sec:results} presents our main results and in Section \ref{sec:conclusions} we draw some conclusions.

\section{The multi-scale network structure of airspace}\label{sec:description}

The airspace can be considered as a multi-scale, dynamic network of interconnected entities. In this section we give a brief overview of the structure of airspaces and we introduce the relevant entities for our analysis. We then describe the different networks that can be defined in the airspace and for each of them we will describe the ``existing partitions" i.e. the network communities already present in the system due to the airspace partitioning in ACCs, NAs and FABs made by Eurocontrol and the national regulatory bodies. Such partitions are primarily driven by the political map of Europe and by operational considerations of the air traffic service providers.

\subsection{Structure of the airspace}

Flights do not currently follow a smooth and optimized trajectory. Instead, they are supposed to follow a path on a predefined mesh, whose nodes are called navigation points, or {\it navpoints}. The position of a navpoint is given by a latitude and a longitude, but not an altitude. A flight plan is therefore a succession of navpoints, together with timestamps and altitudes, that a plane is supposed to follow. However the initial flight plan is not necessarily the one followed during the real trajectory. Due to several disturbances -- such as weather, strikes, congestion, etc -- the actual trajectory might be different from the planned one. It can be a difference in time, altitude, or even in the sequence of navpoints. 

The deviations from the planned trajectory  are typically triggered by the air traffic controller. Specifically the airspace is divided in three dimensional airspace volumes, termed  {\it elementary sector}, or {\it collapsed sector} (called simply {\it sector} in the following). A sector is handled by two controllers: one ``separates''  the aircraft (making sure they do not come too close to each other) in the sector itself, while the other one takes care of the interface with the other sectors. The sectors are dynamic entities, which can be split or gathered depending on the load of traffic. Moreover, the sectors can be roughly divided in two types: the en-route sectors, controlling the planes in their en-route trajectory, and the Terminal Maneuvering Areas (TMA) or the Control Zones (CTR), managing the take-off and landing phases.

The airspaces themselves are bigger, static entities gathering several sectors. The first important one is the {\it Air Control Center} (ACC), where all the sectors are physically controlled from the same room (control center). In the European airspace, called ECAC\footnote{Countries in the enlarged ECAC space are: Iceland (BI), Kosovo (BK), Belgium (EB), Germany-civil (ED), Estonia (EE), Finland (EF), UK (EG), Netherlands (EH), Ireland (EI), Denmark (EK), Luxembourg (EL), Norway (EN), Poland (EP), Sweden (ES), Germany-military (ET), Latvia (EV), Lithuania (EY), Albania (LA), Bulgaria (LB), Cyprus (LC), Croatia (LD), Spain (LE), France (LF), Greece (LG), Hungary (LH), Italy (LI), Slovenia (LJ), Czech Republic (LK), Malta (LM), Monaco (LN), Austria (LO), Portugal (LP), Bosnia-Herzegovina (LQ), Romania (LR), Switzerland (LS), Turkey (LT), Moldova (LU), Macedonia (LW), Gibraltar (LX), Serbia-Montenegro (LY), Slovakia (LZ), Armenia (UD), Georgia (UG), Ukraine (UK).} (European Civil Aviation Conference), there are between $1$ and $5$ ACCs per country. Then we have the {\it National Airspace} (NA), gathering all the ACCs of a single country. The two dimensional boundaries of a NA are very close to the real country's boundaries. On a larger scale, we find the {\it Functional Air Blocks} (FABs) \cite{FABs}, gathering several NAs, like, for example, the Portuguese and Spanish ones. They are not actually operative yet, but they will be important in the so-called new SESAR scenario \cite{sesar}, a future air traffic management scenario that will change dramatically the way air traffic is managed. 

Finally, the last important element are the airports. They act as sinks and sources for the network by ``absorbing'' and ``releasing'' aircraft in the system.

\subsection{Network descriptions of the airspace}

Given the structure summarized above, it is clear that it is possible to define (at least) three different networks describing the airspace. The three networks operate at different spatial and temporal scales, therefore the airspace system can be considered as a multi-scale network. In order to construct the networks, we shall consider a time interval (typically one day) and we define the following graphs. 

The first graph is the {\it network of navigation points}. In this network each node is a navigation point and two nodes are connected if at least one flight goes directly from one node to the other in the considered time interval. Similarly, the second graph is the {\it network of sectors}. Each sector is a node and two nodes are connected if at least one flight goes directly from one node to the other in the considered time interval. Finally the third graph is the \textit{airport network} where nodes are airports and two nodes are connected if at least a flight goes from one node to another in the time interval. 

All the networks are directed and weighted. The weight is given by the number of flights between two nodes in the time interval. As far as the direction is concerned, we notice that most of the graphs are very close to symmetric and therefore one makes a small error in considering the symmetric version of the network, as required by some community detection methods (see below). Finally, note that all these networks are traffic networks. This means that  the links are defined by the traffic in the time interval and are different from a street network where the infrastructure defines the link.

\subsection{Existing partitions of the airspace networks}

The main objective of this paper is the comparison between unsupervised partitions of the different networks of the airspace and the partitions that are already present in the system as a result of its design. We call these partitions, {\it existing partitions}. Here we present the main existing partitions of the three networks that we will use in the following. A set of figures with the different existing partitions for the European airspace is shown in the Appendix.

The navpoint network can be partitioned in terms of national airspaces (NA, see Fig. \ref{navna}) or in terms of control centers (ACC, see Fig. \ref{navacc}).
The  sector network can be partitioned in terms of functional airblocks (FAB, see Fig. \ref{secfab}), in terms of national airspaces (NA, see Fig, \ref{secna}), or in terms of control centers (ACC, see Fig. \ref{secacc}). Finally the airport network can be partitioned in terms of functional airblocks (FAB, see Fig. \ref{airfab}) or in terms of national airspaces (NA, see Fig. \ref{airna}). 

\section{Data}\label{sec:data}

Our database contains detailed information on all the flights that, even partly, cross the ECAC airspace. The data come from two different sources. First, we have access to the Demand Data Repository (DDR) \cite{DDR} from which we have all the trajectories followed by aircraft in the ECAC airspace. In this paper we consider a 28 day time period (termed AIRAC cycle), specifically the one lasting from the $6^{th}$ of May 2010 to the $2^{nd}$ of June 2010. The fact that this cycle is relatively far from the major holidays ensures absence of biases due to seasonalities. A trajectory, called indifferently flight plan here, is made by  a sequence of navigation points crossed by the aircraft, together with altitudes and timestamps. The typical time between two navpoints lies between 1 and 10 minutes, giving a good time resolution for trajectories. In this paper we only use the ``last filed flight plans'', 
which are not the real trajectories flown, but the planned trajectories -- filed from 6 months to one or two hours before the real departure. We do not use the real trajectories because we do not want to include other factors of disturbances, like weather, in our analysis. We selected only scheduled flights -- excluding, in particular, military flights --  using landplanes (regular aircraft) and having a IATA code. This gives, in first approximation, the set of commercial flights. We also excluded all flights having a duration shorter than 10 minutes and a few other flights having obviously data errors.



The other source of information are the NEVAC files, which give the configuration of the airspaces for an entire AIRAC cycle. Here we only used the information on sectors, airspaces and configurations to rebuild the European airspace. Specifically, at each time we have the full three dimensional boundaries of each individual sector and airspace in Europe. 

\section{Methodology}\label{subsec:methods}

\subsection{Community detection methods}



In this article, we consider different algorithms of community detection on networks. Since they use different definitions of communities based on their own method, they give different results and thus allow us to see how robust the obtained partitions are.  Specifically, we consider three methods of community detection:  Infomap \cite{infomap}, Maximization of the modularity with the Louvain method and simulated annealing \cite{Louvain}, and OSLOM \cite{oslom}.

The first algorithm we used is called Infomap \cite{infomap}. The idea behind the method is to consider a random walk over the network. The more the nodes are connected one with each other, the more the walker will stay with them and thus form a community. The analysis of the flows over the network gives access to the underlying community structure. More precisely, the algorithms computes an optimized compressed description of information flows and, from the information theory point of view, the community detection algorithm searches the partition which minimizes the description length of an infinite random walk over the network. This algorithm has a complexity $O(m)$ where $m$ is the number of edges. It is thus efficient with sparse networks, where $m\sim n$, where $n$ is the number of nodes. In our investigation we used the usual implementation of the package, available online \cite{infomap_url}.

The second method is based on the maximization of a function called ``modularity''. 
For a given partition $P$, the modularity $Q$ is the sum of the number of links within each community minus the expected number of links for a given null model, i.e. 
$$
Q=\frac{1}{2m} \sum_{C\in P}\sum_{i,j\in C} \left(A_{ij} - P_{ij}\right),
$$
where $A_{ij}$ is the element $ij$ of the weighted adjacency matrix of the graph, $P_{ij}$ is the element $ij$ of the weighted adjacency matrix under the null model, and $m=\sum_{i,j}A_{ij}/2$ is the total weight in the network. The most popular choice for $P_{ij}$ is the one proposed by Newman and Girvan (NG) \cite{NG}: $P_{ij}^{NG}=k_i k_j/2m$, where $k_i$ is the strength of node $i$. The null hypothesis corresponds to a randomization of the links preserving the strength of each node. It is well known that modularity has a resolution limit \cite{Fortunato07}, i.e. in large networks modularity fails to resolve small communities.

Different choices for the null model in the modularity can also be done. In section \ref{subsec:airports}, we examine a null model which takes into account the spatial localization of the nodes. Following Ref. \cite{spatial}, we use for $P_{ij}$ the following form:
\begin{equation}
P_{ij}^{GEO} = k_i k_j \frac{\sum\limits_{(k,l)|d_{kl}=d_{ij}} A_{kl}}{\sum\limits_{(k,l)|d_{kl}=d_{ij}} k_i k_j},   \label{eq:spa}
\end{equation}
which is the weighted probability for a node to be linked to another node at distance $d$ ($d_{ij}$ being the euclidean distance between nodes $i$ and $j$). In this null model the nodes are more likely to be linked if they are geographically close one to each other. This choice allows to see the communities which are not only explained by their geographical proximity.

Different computational methods can be used to find the maximum of modularity. One of the most popular is the Louvain method, an algorithm which computes the communities, then the induced graph -- i.e. where each node is a community -- then the communities of this graph, until a maximum in modularity is reached. This method is very efficient since the complexity is $O(m)$,  and it gives accurate results. We used the software package available at \cite{Louvain_url}. However, this cannot be used straightforwardly when the null model is the one of Eq. \ref{eq:spa}. Since the probability needs explicitly some geographical coordinates, computing the induced graph is meaningless, because one cannot associate each node (community) with spatial coordinates. 


Instead, when using the null model of equation Eq. \ref{eq:spa}, we choose a simulated annealing method. The simulated annealing, based on a physical process used to change the properties of glass or metals in the industry, is usually used to find a minimum (or maximum) of a non monotonic multi-dimensional function. It is based on a random walk in the phase space: at each step, one changes slightly the system (here for instance, changing the assignment of a node to a community) and see if the function to be optimized (the ``energy'') has decreased or increased. Given that we look for a minimum here, we always accept the new state in the first case (decrease), and randomly choose if we accept the new state in the latter case (increase). The most widely chosen probability of acceptance is usually of the form $e^{-\frac{\delta E}{T}}$, where $\delta E$ is the change in the function between the new and the previous state, and $T$ is a parameter called temperature, in analogy with the physical process. The simulated annealing algorithm itself consists in a progressive decrease of $T$, so that the system first explores big wells of energy, then progressively gets trapped in deeper, narrower wells. The algorithm gives accurate results but needs much more time to converge than the previous methods. As a final comment, when applying the modularity partition (with the Louvain method or with the simulated annealing) we considered for technical reasons the undirected version of the networks. As mentioned above, air traffic networks are highly symmetric and therefore the error should be small. 

In the case of the navpoint and sector network, we used the Louvain method, considering the NG choice $P_{ij}^{NG}=k_i k_j/2m$. In the case of the airport network we used both the NG prescription and the one of Eq. $(\ref{eq:spa})$ that takes into account the geographical constraints, hence using both the Louvain method and the simulated annealing.

The last method of community detection we use is called OSLOM -- for Order Statistics Local Optimization Method \cite{oslom}. Its general principle is the following. Using the NG null model presented in the standard definition of modularity, the authors use a fitness function -- based on the probability that an external node to a community has a given number of neighbors within this community --  to assess the statistical significance of each community. More precisely, each external nodes is ranked following this fitness function, and the algorithm tests the likelihood of the score, with the given rank of this node against the null model. This procedure has several benefits. Since this optimization is local, i.e. made independently for each community, the result can be a partition with overlapping clusters. Moreover, the method can be used as refinements to other methods (Infomap, modularity), because one can give to the algorithm an existing partition as input. It allows also to use the efficiency of the other detection techniques. The method's complexity itself cannot be exactly computed, but numerical results shows that it is close to $O(n)$.  The package is available at \cite{oslom_url}.

\subsection{Metrics for comparison of partitions} \label{subsec:metrics}

In order to compare the partitions given by the different algorithms, 
we used two different metrics. 

The first one is called the Rand index (RI) \cite{Rand71} and is computed in the following way. Given two different partitions $P_1$ and $P_2$ of the same set and a pair of elements in this set, there can be four cases: a) the elements are in the same community in $P_1$ and $P_2$, b) they are in different communities in $P_1$ and $P_2$, c) they are in the same community in $P_1$ but in different ones in $P_2$, and d) they are in the same in $P_2$ but not in $P_1$. The Rand index is simply the number of occurrences of the two first cases a and b over the total number of possibilities. The Rand index ranges between $0$ and $1$

We introduce also the asymmetric Rand index (ARI) \cite{Hubert85} to have a measure of inclusion relationships. A partition $P_1$ will be said to be included in a partition $P_2$ if each cluster of $P_1$ is included in a cluster of $P_2$. The asymmetric Rand index is computed as the ratio of the first three cases a, b, c, above over the total number of cases. Of course, the ARI is no longer symmetric and it reaches 1 if the partition $P_1$ is included in $P_2$.


The second metric is the  mutual information (MI). The mutual information  between two random variables $X$ and $Y$ is
$$
I(X,Y) = \sum_{y}\sum_x p(x,y) \log \left(\frac{p(x,y)}{p(x)p(y)}\right),
$$
where $p(x)$ and $p(y)$ are the marginal probabilities to have $X=x$ and $Y=y$ and $p(x,y)$ is the joint probability. The mutual information is symmetric and is related to the joint and conditional entropies of $X$ and $Y$. This definition can be used to compare partitions, where $X$ and $Y$ represent the labels of communities in each partition.

Finally, note that we used normalized and adjusted versions of these two metrics. The latter term means that each metric takes into account the probability to have a non-zero value for two purely random realizations of $X$ and $Y$, due to the finite size of sample. Hence, after being adjusted and normalized,  RI lies between -1 and 1 and MI lies between 0 and 1. The value $0$ implies that the two partitions have no more in common than two random partitions would have. The value $1$ implies that the two partitions are exactly the same. 

\subsection{Community characterization} 

After a network has been partitioned into communities, the next step is to characterize the nodes composing each community. Characterization of a community means to identify which characteristics of the nodes is ``overexpressed" \cite{hyper}. When we will consider the partition of the network of airports (see section \ref{subsec:airports}) we will give to each node (airport) a label indicating the country the airport belong to. Then we will ask the question of which countries, if any, are overexpressed in each community.

In order to do it, let us consider a network where the nodes have all a label $L$ which can take several values $\{L_1, ..., L_n\}$. For each label $L_i$, we want to see if it is overexpressed in a given cluster $C$. The probability of having $N_{L_i}^C$ under the null hypothesis that elements in the community are randomly selected is given by the hypergeometric distribution \cite{hyper}, i.e. 
$$
P(N_{L_i}^C|N^C, N_{L_i}, N)=\frac{{{N^C}\choose{N^C_{L_i}}} {{N-N^C}\choose{N_{L_i} - N^C_{L_i}}}}{{{N}\choose{N_{L_i}}}}
$$
where $N_{L_i}^C$ is the number of nodes within $C$ with label $L_i$, $N^C$ is the total number of nodes in $C$, $N_{L_i}$ is the total number of nodes in the network having label $L_i$, and $N$ is the total number of nodes. Then the probability that $C$ has more than $N_{L_i}^C$ nodes with label $L_i$ is given by $p(N_{L_i}^C)=1-\sum_{k=0}^{N_{L_i}^C} P(k|N^C, N_{L_i}, N)$. If this probability is below a given threshold $p$, then $L_i$ is said to be overexpressed in $C$, because it is unlikely to find such a number of nodes in this cluster with this label only by chance. Since we are performing several tests on the different values of $L$ and on different clusters $C$, the value of $p$ has to be tuned in order to avoid the false positives, increasing when tests are used repeatedly. One way of dealing with multiple hypothesis tests is to set the threshold to $p/N_t$, where $N_t$ is the number of tests. This correction, known as the Bonferroni correction, is the most conservative, but other corrections can be found in the literature \cite{Miller}.


\section{Results}\label{sec:results}

\subsection{The navigation point network}

The navigation point network is the most detailed network we consider in this paper. To the best of our knowledge, this type of network has been considered only  in Ref. \cite{Cai12}, where some basic network metrics for the Chinese airspace have been studied. Before discussing the community detection result, we present some network metrics for the European case.

\subsubsection{Network metrics}

The navpoints are defined with a longitude and a latitude, but without an altitude. As a consequence they can be used conjointly by several sectors at different altitude. Thus, a navpoint cannot be considered as a single entity/node in terms of the network. On the other hand, the structure of the airspace has an interesting property which is of use here. Indeed, three phases are usually defined for trajectories of flights: the take-off, the en-route, and the landing phases. If one considers only the en-route part, the corresponding portion of airspace becomes much simpler. At high altitudes, navpoints are only crossing one or two sectors vertically, and these sectors often have the same 2d boundaries. Thus, we chose to consider only the upper part of all the trajectories, cutting them below  $24,000$ feet -- a value based on operational considerations. This way, navpoints are contained in only one or two sectors and are meaningful entities.

\begin{figure}[htbp]
\centering
\includegraphics[width=0.48\textwidth]{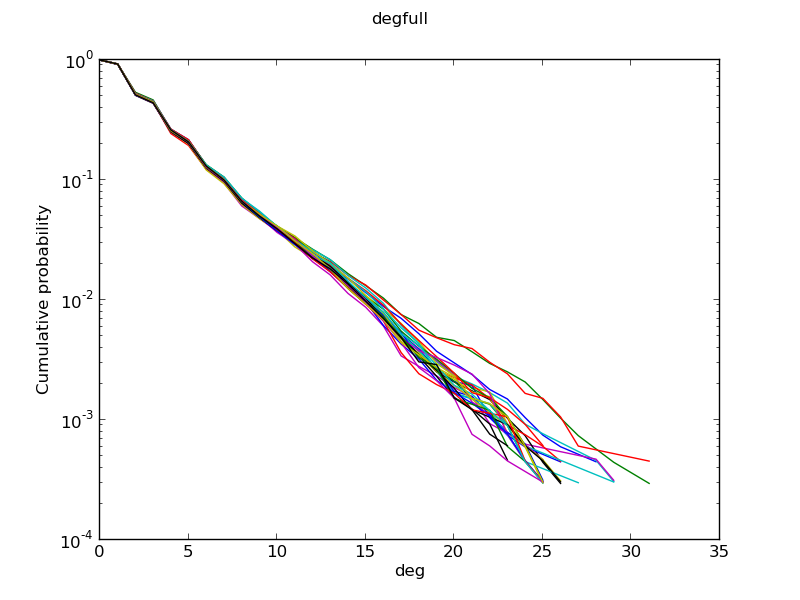}
\includegraphics[width=0.48\textwidth]{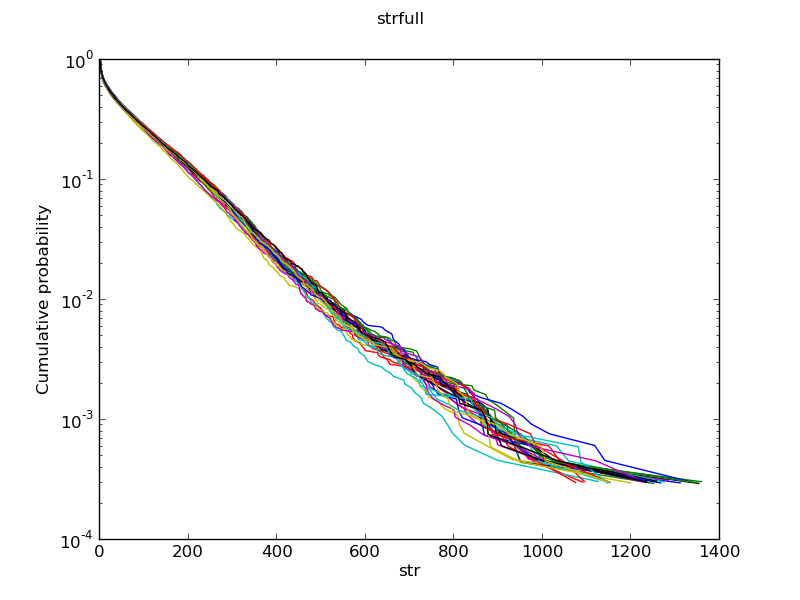}
\caption{Distribution of degree (left) and strength (right) in the navigation point network in a semilog scale. Each color corresponds to a different day of the AIRAC cycle. The mean degree is 3.88 with standard deviation 3.0, with min and max respectively equal to 1 and 29. The mean strength is 87.0 with standard deviation 118.7 and min/max equal to 1 and 1338.}
\label{fig:distributions_navpoints}
\end{figure}

Based on these trajectories, we build the navpoint network. Each of them is a node and a link exists if at least one flight goes from one to the other on the given time-frame (one day). The link is weighted with the number of flights. For the whole ECAC space, the network has around $6,000$ nodes, and the $21,000$ flights per day create $12,000$ edges. The distribution of degree and strength of the navpoint network are presented in Fig. \ref{fig:distributions_navpoints}. The distribution is quite constant in time and is very close to an exponential. This is an expected behavior, since this network is strongly constrained geographically. Indeed, an aircraft cannot skip many navigation points to go from one point to another, and is bound to travel through geographical neighboring points.

\subsubsection{Communities}

We  now consider the clusters formed by the traffic on the navpoint network. Specifically,
 our aim is to see if the bottom-up clusterizations made by the algorithms are consistent with the top-down design which lead to their creation. Indeed, the ACCs are designed by partitioning a certain given volume (the national airspace) and the sectors are partitioning these pieces. Even though the details of the partitions depend obviously on the traffic demand, they are not at the root of the design. Instead, by using unsupervised clusterization, we do not have any prior on the partition, other than the boundaries of the ECAC space. In this sense, this approach is complementary to the existing top-down ``expert'' partitioning and can give insights on how to improve it.

We use the different methods presented in section \ref{subsec:methods} to generate the partitions and compare them with the tools presented in section \ref{subsec:metrics}. Moreover, we build the partitions based on the nationalities and the ACCs  (see Appendix).

\begin{figure}[t]
\centering
\includegraphics[width=0.75\textwidth]{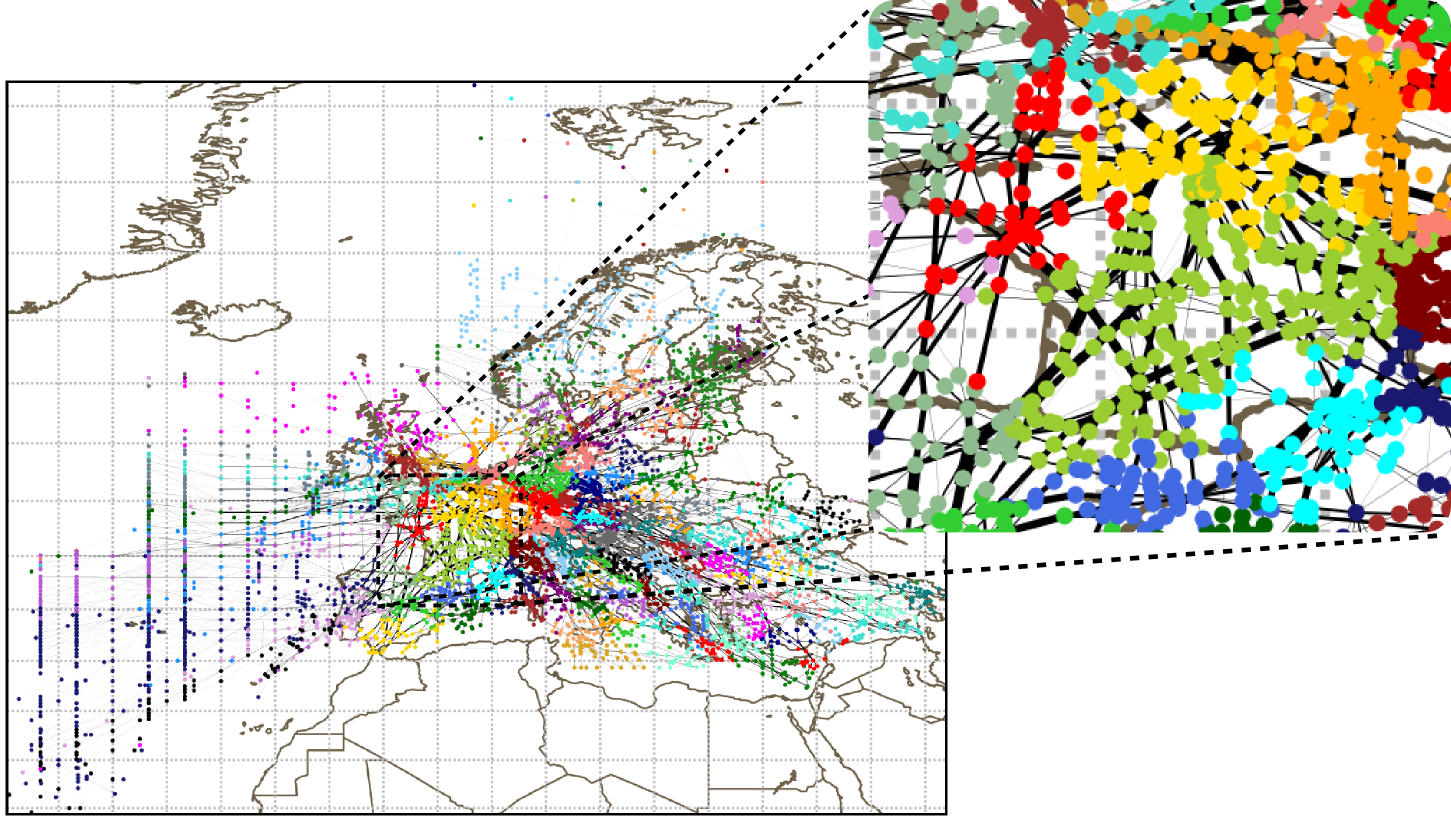}
\caption{Communities obtained with the OSLOM algorithm on the network of navpoints for May 6, 2010. Each color corresponds to a different community. }
\label{fig:map_oslom_navpoints}
\end{figure}

Figure \ref{fig:map_oslom_navpoints} shows one of these partitions, obtained with the OSLOM algorithm. The clusters are clearly geographical, with almost no geometrical overlap between them. For this algorithm, communities are much smaller than the national scale, but their boundaries seem roughly consistent with the national boundaries. 

We perform an extensive community detection by using the three algorithms described in Section \ref{subsec:methods} for each of the $28$ days of the AIRAC cycle.  In the case of the modularity-based algorithms we consider $P_{NG}=k_i k_j/ 2 m$.  The average number of communities for each algorithm is reported in Table \ref{tab:nc_navpoints}. We notice that the numbers of communities depends significantly on the adopted algorithm. The coarsest level of description is given by the modularity maximization, with only $42.8$ clusters, even less than the number of countries. OSLOM gives more than three times this number ($150.0$), and twice the number of ACCs ($88.5$). Finally, Infomap gives around $600$ communities, a very large number, comparable only to the number of sectors.

In the second line of Table \ref{tab:nc_navpoints} we show the minimum number of communities (induced or existing) which involve 90\% of the nodes in the network. These figures indicate that a fraction of the communities ranging from 30\% (ACC) to 50\% (OSLOM) contains only 10\% of the nodes. These are therefore very small communities. It is surprising to see that such small communities are present not only in the induced partitions but also in the existing ones.

\begin{table}[htbp]
\begin{tabular}{|c|c|c|c|c|c|}
\hline
Partition 	& Infomap			& Modularity	& OSLOM			& ACC 			& NA\\
\hline
$N_c$		& 598.0$\pm$14.7	& 42.8$\pm$3.9	&150.0$\pm$21.6	& 75.5$\pm$2.6	& 46.0$\pm$0.\\
\hline		
$N_c^{0.9}$	& 408.8$\pm$8.5		& 25.9$\pm$1.6	& 75.1$\pm$4.4	& 53.3$\pm$2.2	& 20.0$\pm$0.\\
\hline
\end{tabular}
\caption{Average number of communities in the network of navpoints and average minimum number of communities containing 90\% of the nodes. Computed over 28 days, with standard deviations taken as error bars.}
\label{tab:nc_navpoints}
\end{table}

To refine these comparisons, we use the different metrics described in Section \ref{subsec:metrics}. The partitions have been computed over the $28$ days of the AIRAC cycle and the MI and RI averaged on this period. The idea is to use these metrics to understand how the community detection results are similar with each other and how they compare with the existing partitions of ACCs, NAs and FABs illustrated in the appendix. The results of this systematic comparison between partitions are given in Tables \ref{tab:navpoint_rand}, \ref{tab:navpoint_nmi}, and \ref{tab:navpoint_randA}. 

Focusing first on the Rand index (Table \ref{tab:navpoint_rand}), we see that the Infomap partition has little in common with the others (at least when one uses the Rand index). This result is somehow expected. In fact, the Infomap communities are many more and  smaller than those obtained with the other two algorithms. The other two induced partitions are roughly equally close to each other and to the existing partitions. More precisely, OSLOM is closer to the ACCs partition and modularity to the national partition, as expected from the number of communities. 
\begin{table}[htbp]
\centering
\begin{tabular}{|c||c|c|c|c|c|}
\hline
\bf{Rand Index} & Infomap			& Modularity			& OSLOM 				& ACC				& NA\\
\hline
\hline
Infomap		& 1.				&0.074$\pm$0.007	& 0.14 $\pm$0.01		&0.078$\pm$0.003	& 0.043$\pm$0.002\\
\hline
Modularity	&0.074$\pm$0.007	& 1.				&0.27$\pm$0.03			&0.28$\pm$0.02	& 0.30$\pm$0.03\\
\hline
OSLOM		&0.14 $\pm$0.01		&0.27$\pm$0.03		&1.						&0.28$\pm$0.02	& 0.23$\pm$0.02\\
\hline	
ACC			&0.078$\pm$0.003	&0.28$\pm$0.02		&0.28$\pm$0.02		&1.					& 0.68$\pm$0.01\\
\hline
NA 			& 0.043$\pm$0.002	&  0.30$\pm$0.03	&  0.23$\pm$0.02		& 0.68$\pm$0.01	&1.\\
\hline
\end{tabular}
\caption{Comparisons of the partitions using the Rand index. The numbers are the average Rand index between the partitions and the error bars are standard deviations.}
\label{tab:navpoint_rand}
\end{table}

The mutual information (see Table \ref{tab:navpoint_nmi}) confirms these results, although the score for the Infomap is much higher, compared to the Rand index. The modularity partition is now equally close to the ACC and NA existing partitions.
\begin{table}[htbp]
\centering
\begin{tabular}{|c||c|c|c|c|c|c|}
\hline
\bf{Mutual Information} & Infomap			& Modularity	& OSLOM 			& ACC			& NA\\
\hline
\hline
Infomap					& 1.				&0.38$\pm$0.008	& 0.42 $\pm$0.008	&0.37$\pm$0.004	& 0.32$\pm$0.003\\
\hline
Modularity				&0.38$\pm$0.008		& 1.			&0.54$\pm$0.01		&0.57$\pm$0.01	& 0.57$\pm$0.01\\
\hline
OSLOM					& 0.42 $\pm$0.008	&0.54$\pm$0.01	&1.					&0.59$\pm$0.009	& 0.51$\pm$0.09\\
\hline
ACC						&0.37$\pm$0.004		&0.57$\pm$0.01	&0.59$\pm$0.009		&1.				& 0.80$\pm$0.09\\
\hline
NA 						&0.32$\pm$0.003		& 0.57$\pm$0.01	& 0.51$\pm$0.09		& 0.80$\pm$0.09	&1.\\
\hline
\end{tabular}
\caption{Comparisons of the partitions using the mutual information. The numbers are the average mutual information between the partitions and the error bars are standard deviations.}
\label{tab:navpoint_nmi}
\end{table}

The  Table \ref{tab:navpoint_randA} presents the asymmetric Rand index, as defined in \ref{subsec:metrics}. It gives additional information on the inclusion relation between partitions. For instance, values which are almost symmetric and close to the corresponding ones in Table \ref{tab:navpoint_rand} show that the given partitions do not have a inclusion relationship between them. On the contrary, highly asymmetric values indicate that a partition is included in another one. The first column of Table \ref{tab:navpoint_randA} displays some values much higher than the corresponding ones in Table \ref{tab:navpoint_rand}. This means that Infomap gives very small clusters, which are more likely to be included in bigger communities obtained from other partitions. The inclusion of Infomap partition into the modularity partition is quite interesting. In fact, the data indicates that this inclusion is more significant than the one of  the Infomap partition into the NA partition. This means that the Infomap partition can be viewed as a ``refinement'' of the modularity partition. This last  partition includes also quite well the OSLOM one, at least much better than the ACCs. It is also quite ``transversal'' to the NAs.  In fact their symmetric score is very low, given their similarity in the sizes of the communities, and  the asymmetric scores are close to each other, showing that there is no strong relationship of inclusion between them.
\begin{table}[htbp]
\centering
\begin{tabular}{|c||c|c|c|c|c|}
\hline
\bf{Asymmetric Rand Index} 	& Infomap			& Modularity		& OSLOM 				& ACC				& NA\\
\hline
\hline
Infomap						& 1.				&0.039$\pm$0.004	& 0.080$\pm$0.086		&0.038$\pm$0.001	& 0.022$\pm$0.0008\\
\hline
Modularity					&0.82$\pm$0.01		& 1.				&0.53$\pm$0.05			&0.33$\pm$0.03		& 0.25$\pm$0.03\\
\hline
OSLOM						&0.57 $\pm$0.01		&0.18$\pm$0.02		&1.						&0.20$\pm$0.02		& 0.14$\pm$0.02\\
\hline
ACC							&0.62$\pm$0.01		&0.28$\pm$0.02		&0.25$\pm$0.02			&1.					& 0.51$\pm$0.01\\
\hline
NA 							&0.70$\pm$0.001		& 0.37$\pm$0.03		& 0.62$\pm$0.04			& 1.				&1.\\
\hline
\end{tabular}
\caption{Comparisons of the partitions using the asymmetric Rand index. The values in this table are not necessarily symmetric, because they take into account the inclusion relationships. A high score means that the partition of the given column is included in the partition of the given line. The numbers are the average asymmetric Rand index between the partitions and the error bars are standard deviations.}
\label{tab:navpoint_randA}
\end{table}

Surprisingly, the OSLOM partition has a high number of communities and nevertheless scores similar to the modularity partition when both are compared with ACCs and NAs. This suggests that indeed the modularity partition somehow reproduces the organization of airspace in NAs, while the OSLOM partition reproduces the organization in ACCs, which are a sub-partition of the NAs. In this respect the OSLOM partition refines the modularity one at a smaller scale.
As to the Infomap, its very small communities seem a hierarchical partition of the modularity partition, as indicated by the asymmetric Rand index, with a typical size similar to the sector level. 

All these results underline the difference from the bottom-up approach with an unsupervised partition based on the traffic and the top-down approach of the real construction of the existing partitions. The next section deals with the same idea at a coarser level: given some predefined partitions of the navpoints (sectors), we ask how these clusters can be gathered based on the traffic from one sector to the other.

\subsection{The sector network}

	
The sectors are the smallest operational pieces of airspace and as such are controlled by a couple of controllers. One in particular is in charge of the interaction with the other neighboring sectors. Thus, building the network of sectors, where nodes are sectors and links are built from traffic data between two sectors gives the finest description of the operative European airspace. To the best of our knowledge, this network has never been studied in the literature. As for the navpoint network,  we present first some basic network metrics and then we move to the communities detection problem. Here, we used the full network of sectors, without cutting the trajectories in altitude, and without discarding the airports areas.

\subsubsection{Classic metrics}

\begin{figure}[t]
\centering
\includegraphics[width=0.6\textwidth]{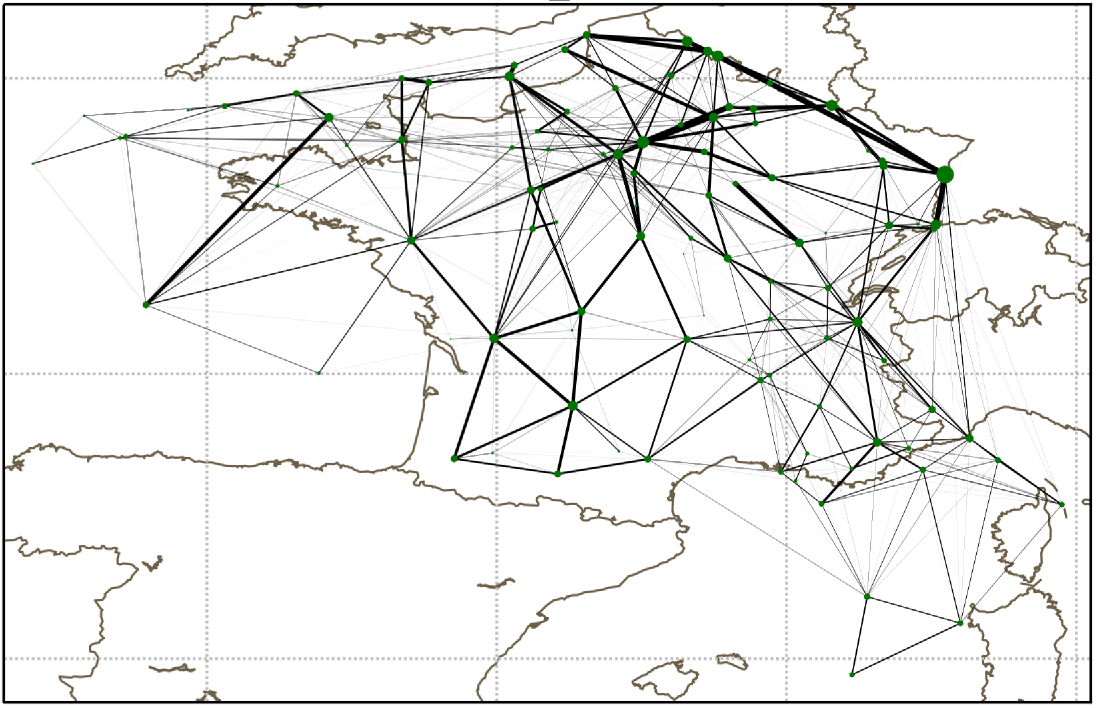}
\caption{Example of the sector network for the French airspace. Each node represents a sector, the thickness of the links between them is proportional to the number of flights passing directly from one sector to the other. Sectors which are exactly on top of each other are a bit shifted to see them.
}
\label{fig:LF_sectors}
\end{figure}

The topology of the sector network is dynamic, because sectors can be split or collapsed depending on the expected and real demand. The number of active sectors in Europe varies from $300$ (around midnight) to more than $700$ during the day. In order to build the network, we considered that there is a link between two nodes if a flight goes directly from one sector to the other in the considered time window. The links are also weighted the same way, i.e. by using the number of flights. 

\begin{figure}[hbtp]
\centering
\includegraphics[width=0.48\textwidth]{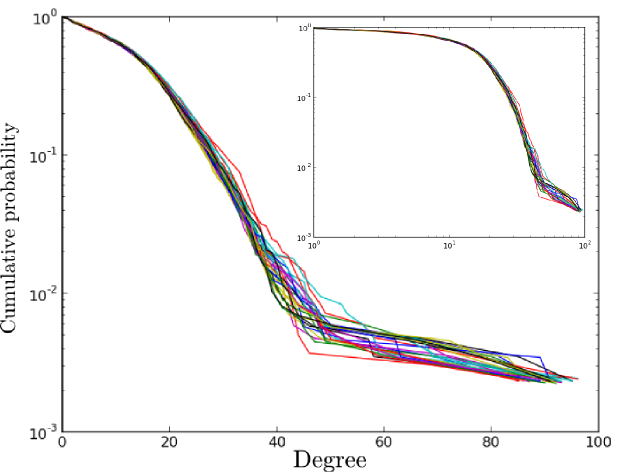}
\includegraphics[width=0.48\textwidth]{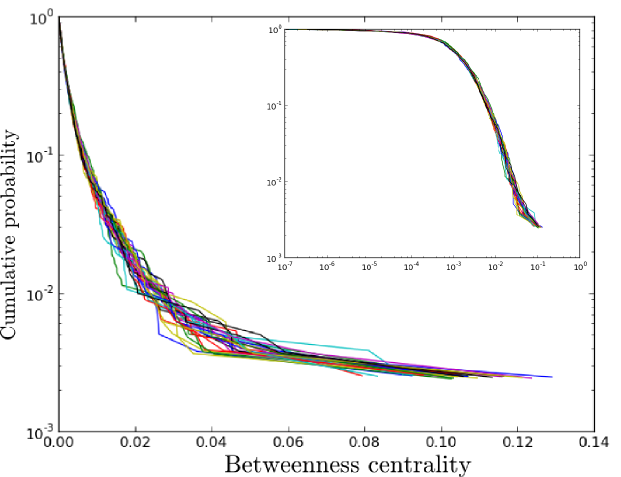}
\caption{Distribution of degree (left) and betweenness centrality (right) of the sector network of the European airspace in semi-log scale (insets: in log-log scale). Each color corresponds to a different day of the AIRAC cycles. The mean degree of the network averaged on $28$ days is $11.4$, with a standard deviation equal to $0.3$. The intraday standard deviation is however very large, being equal to $11.5\pm0.2$. The network is thus heterogeneous, even if not scale free. The corresponding values for strength are $194.3\pm9.4$ for the mean and $356.5\pm14.8$ for the deviation in one day. Finally the betweenness centrality displays a averaged mean degree of $2.0~10^{-3}\pm0.11~10^{-4}$ and a deviation of $7.2\pm0.43$.}
\label{fig:distributions_sectors}
\end{figure}

An example of the sector  network of the French airspace is given in Fig. \ref{fig:LF_sectors}. For the whole Europe, the network has around $6,500$ links per day. The degree distribution is displayed in the left panel of Fig. \ref{fig:distributions_sectors}. As one can see, the distribution displayed is distinct from a power law, but also from an exponential. We notice here the presence of very few hubs (mostly German sectors), i.e. central sectors which redistribute the traffic around Germany in many other sectors. The strength distribution (not displayed here, but available upon request) is a bit closer to an exponential. The betweenness centrality, shown on the right panel of Fig. \ref{fig:distributions_sectors}, does not seem to be a power law nor an exponential. In fact, the network is far from being homogeneous and the standard deviations computed within a day are of the same order or greater than the average degree, strength, or betweenness, which means that the distributions are very wide.  


These characteristics and the difference from a power law distribution are the results of geographical constraints. Indeed, even if the network is not exactly planar, there is no ``short-cut'' between far away sectors. An aircraft has to go through a sequence of nodes to reach the destination. This implies a node is more likely to be connected to its geographical neighbors, which are of course of limited number.

\subsubsection{Communities}

We use the same methods described earlier for the detection of communities in the network of sectors. Moreover, we use three existing partitions: the ACCs, the NAs, and the FABs (see maps of figures  \ref{secfab}, \ref{secna}, and \ref{secacc} in the Appendix for these partitions).

Figure \ref{fig:ECAC_sectors} shows the result of the Infomap method on this network. The detected communities are typically much smaller than the typical size of a country. However it is interesting to notice that they are not transnational either, and thus they seem to partition the national airspaces themselves. Moreover, some known specificities of the European airspace are well recovered. For example, the two big communities of Ireland and North United Kingdom for instance are present, as well as the four ACCs of Italy.

\begin{figure}[htbp]
\centering
\includegraphics[width=0.7\textwidth]{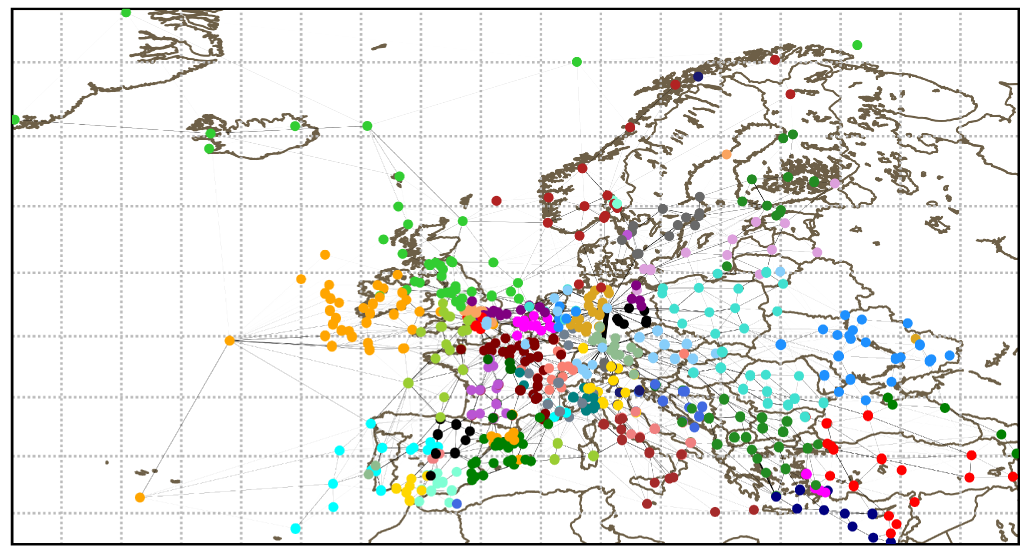}
\caption{Community detection in the European network of sectors using the Infomap algorithm. The network is computed on the May 6, 2010. Each color corresponds to a different community.}
\label{fig:ECAC_sectors}
\end{figure}

Table \ref{tab:nc_sectors} gives the number of communities for each partition, averaged across the days. Since the number of communities is related to their typical size, we notice that the average size of the communities are quite different from one algorithm to the other. The modularity  method, due probably to resolution effects (see \ref{subsec:methods}), gives the biggest ones. On the contrary, the OSLOM and Infomap give smaller clusters of the typical size of the country. These considerations are clearly not sufficient though because, as we can see from figure \ref{fig:ECAC_sectors}, the community sizes are quite heterogeneous over Europe.

In the second line of Table \ref{tab:nc_sectors} we show the minimum number of communities (induced or existing) that involve 90\% of the nodes in the network.  As for the navpoint network,  also here there are very small communities. However, the number of small communities is here smaller than in the previous case. This might be an indication of the fact that sectors are more inter-connected than navpoints.
The only obvious exception is given by the NA case, where now almost 50\% of communities contains 10\% of nodes while such percentage was 30\% in the case of the navpoint network. This is clearly due to a high number of small countries such as Belgium, Netherlands, etc.

\begin{table}
\begin{tabular}{|c|c|c|c|c|c|c|}
\hline
Partition 	& Infomap		& Modularity	& OSLOM			& ACC 			& NA 			& FABs\\
\hline
$N_c$		& 66.7$\pm$3.3	& 12.6$\pm$1.1	& 49.9$\pm$3.0	& 115.4$\pm$2.0	& 49.5$\pm$0.7	& 12.0$\pm$0.\\
\hline
$N_c^{0.9}$	& 42.9$\pm$3.0	& 11.1$\pm$0.7	& 39.5$\pm$2.5	& 58.7$\pm$0.8	& 24.0$\pm$0.	& 9.0$\pm$0.\\
\hline
\end{tabular}
\caption{Average number of communities in the network of sectors and average minimum number of communities containing 90\% of the nodes. Computed over 28 days, with standard deviations taken as error bars. The number of ACCs is not stable because some of them are not used during the period.}
\label{tab:nc_sectors}
\end{table}

A more advanced comparison between all these partitions is presented in Tables \ref{tab:compare_sectors_rand}, \ref{tab:compare_sectors_nmi}, and \ref{tab:compare_sectors_rand_a}. 
In the first two, displaying the Rand index and the mutual information, we see that the 
Infomap and OSLOM methods in particular are providing partitions which are quite close to the ACCs. Another interesting point is that the modularity method, despite having a similar average size of communities, is not close to the FABs partition. Therefore, it is worth emphasizing that their boundaries are very different from the FABs, and this suggests a possible alternative design to the FABs which preserves approximately the total number of communities.
\begin{table}
\centering
\begin{tabular}{|c||c|c|c|c|c|c|}
\hline
\bf{Rand Index} & Infomap			& Modularity		& OSLOM 			& ACC 				& NA			& FABs\\
\hline
\hline
Infomap			& 1.				&0.28$\pm$0.03		& 0.44 $\pm$0.04	&0.38$\pm$0.03		&0.27$\pm$0.02	& 0.17$\pm$0.01\\
\hline
Modularity		&0.28$\pm$0.03		& 1.				&0.26$\pm$0.02		&0.21$\pm$0.02		&0.37$\pm$0.03	& 0.32$\pm$0.03\\
\hline			
OSLOM			&0.44$\pm$0.04		&0.26$\pm$0.02		&1.					&0.42$\pm$0.03		&0.32$\pm$0.03	& 0.20$\pm$0.02\\
\hline	
ACC 			&0.38$\pm$0.03		&0.21$\pm$0.02		&0.42$\pm$0.03		&1.					&0.48$\pm$0.02	&0.24$\pm$0.008 \\
\hline
NA				&0.27$\pm$0.02		&0.37$\pm$0.03		&0.32$\pm$0.03		&0.48$\pm$0.02		&1.				&0.6$\pm$0.006 \\
\hline
FABs 			&0.17$\pm$0.01		&0.32$\pm$0.03		&0.20$\pm$0.02		& 0.24$\pm$0.008	&0.6$\pm$0.006	&1.\\
\hline
\end{tabular}
\caption{\label{tab:compare_sectors_rand} Comparisons between partitions for the network of sectors using the Rand index. The values are averaged over 28 days and the error bars displayed are the standard deviations.}
\end{table}

\begin{table}
\centering
\begin{tabular}{|c||c|c|c|c|c|c|}
\hline
\bf{Mutual Information} & Infomap			& Modularity		& OSLOM 			& ACC 				& NA			& FABs\\
\hline
\hline
Infomap					& 1.				&0.47$\pm$0.01		& 0.64 $\pm$0.02	&0.58$\pm$0.1		&0.52$\pm$0.01	& 0.40$\pm$0.007\\
\hline
Modularity				&0.47$\pm$0.01		& 1.				&0.47$\pm$0.01		&0.39$\pm$0.01		&0.53$\pm$0.02 	& 0.54$\pm$0.02\\
\hline
OSLOM					&0.64 $\pm$0.02		&0.47$\pm$0.01		&1.					&0.58$\pm$0.01		&0.55$\pm$0.02 	& 0.43$\pm$0.01\\
\hline	
ACC 					&0.58$\pm$0.01		&0.39$\pm$0.01		&0.58$\pm$0.01		&1.					&0.66$\pm$0.05	&0.45$\pm$0.002\\
\hline
NA						&0.52$\pm$0.01		&0.53$\pm$0.02		&0.55$\pm$0.02		&0.66$\pm$0.05		&1.				& 0.67$\pm$0.002 \\
\hline
FABs 					&0.40$\pm$0.007		&0.43$\pm$0.01		&0.43$\pm$0.01		&0.45$\pm$0.002		&0.67$\pm$0.002 & 1.\\
\hline
\end{tabular}
\caption{Comparisons between partitions for the network of sectors using the Mutual Information. The values are averaged over 28 days and the error bars displayed are the standard deviations.}
\label{tab:compare_sectors_nmi}
\end{table}

In Table \ref{tab:compare_sectors_rand_a}, displaying the asymmetric Rand index, we see that the OSLOM and Infomap partitions have symmetric values, on the contrary of the modularity method with the two others. Thus, the latter seems to be a coarser version of the two others, even though the inclusion is far from perfect. The same remark holds for the ACCs, which are not perfectly included in the modularity partition. This is expected since we have seen that this partition has boundaries very different from the national ones. Finally, we note that the FABs are also not perfectly including the Infomap and OSLOM partition, despite their much bigger size.
\\

\begin{table}
\centering
\begin{tabular}{|c||c|c|c|c|c|c|}
\hline
\bf{Asymmetric Rand Index} 	& Infomap			& Modularity		& OSLOM 			& ACC 				& NA			& FABs\\
\hline
\hline
Infomap						& 1.				&0.17$\pm$0.02		& 0.43$\pm$0.04		&0.39$\pm$0.03		&0.18$\pm$0.01 	& 0.10$\pm$0.008\\
\hline
Modularity					&0.74$\pm$0.04		& 1.				&0.67$\pm$0.03		&0.60$\pm$0.04		&0.45$\pm$0.04	 & 0.26$\pm$0.03\\
\hline
OSLOM						&0.44 $\pm$0.04		&0.16$\pm$0.02		&1.					&0.51$\pm$0.05		&0.22$\pm$0.02	 & 0.11$\pm$0.01\\
\hline	
ACC 						&0.37$\pm$0.04		&0.13$\pm$0.01		&0.40$\pm$0.03		&1.					&0.31$\pm$0.01	 &0.13$\pm$0.005\\
\hline	
NA							&0.55$\pm$0.04		&0.32$\pm$0.03		&0.63$\pm$0.03		&1.					&1.				 & 0.43$\pm$0.006\\
\hline	
FABs 						&0.69$\pm$0.05		&0.43$\pm$0.05		&0.76$\pm$0.04		&1					&1				 & 1.\\
\hline	
\end{tabular}
\caption{Comparisons between partitions for the network of sectors using the asymmetric Rand index. The values are averaged over 28 days and the error bars displayed are the standard deviations.}
\label{tab:compare_sectors_rand_a}
\end{table}

In conclusion, the partitions inferred by the different methods, although relatively close to the existing partitions of the European airspace, are distinct from them. This means that a new design for the European airspace based on these unsupervised detected partitions could be more optimized because the new ACCs would be more densely connected inside and have less interface (links) with the adjacent ones. It could also help devising dedicated coordination tools and procedures by identifying the boundaries with high traffic exchange volumes.




\subsection{The airport network} \label{subsec:airports}

The last network we consider is the airport network. As explained in the Introduction, this is probably the most studied air traffic network, also for its relation with socio-economical phenomena, such as passenger mobility and epidemic spreading.  For this reason we do not present a detailed analysis of the network metrics. In accordance with previous studies \cite{LiPing03,Barrat04,Li04,Guimera05,Barrat05, Guida07,Bagler08,Xu08,Quartieri08,Zanin08,Xu11,Wang11,Lillo11} we found that the distributions of degree and strength has a power law tail. This reveals the presence of hubs, i.e. nodes with a high degree and a high strength which shorten significantly the paths on the network (in terms of number of nodes). This feature is also revealed when one studies the relationship between the betweenness centrality and the degree. A summary of the empirical analysis of the airport network obtained from the data investigated here can be found in Ref. \cite{Lillo11}.

If the network metrics of the airport network have been extensively studied, to the best of our knowledge there are no studies considering the community structure of this network. Following the previous analysis on sectors and navpoints, we study here the relationship between this structure and operating division of airspace. We will see that the interpretation of the communities in the airport network is somehow different from the previous one concerning sectors and napvoints.

\subsubsection{Communities}

We consider the network of airports. In this graph, airports are nodes and a link between them exists if the two nodes are connected (in the investigated time window) by one or more flights. The link is then weighted according to the number of flights between the nodes. Here the European network has between $480$ and $510$ nodes, depending on the day, and around $6,300$ directed edges.

\begin{figure}[htbp]
\centering
\includegraphics[width=0.6\textwidth]{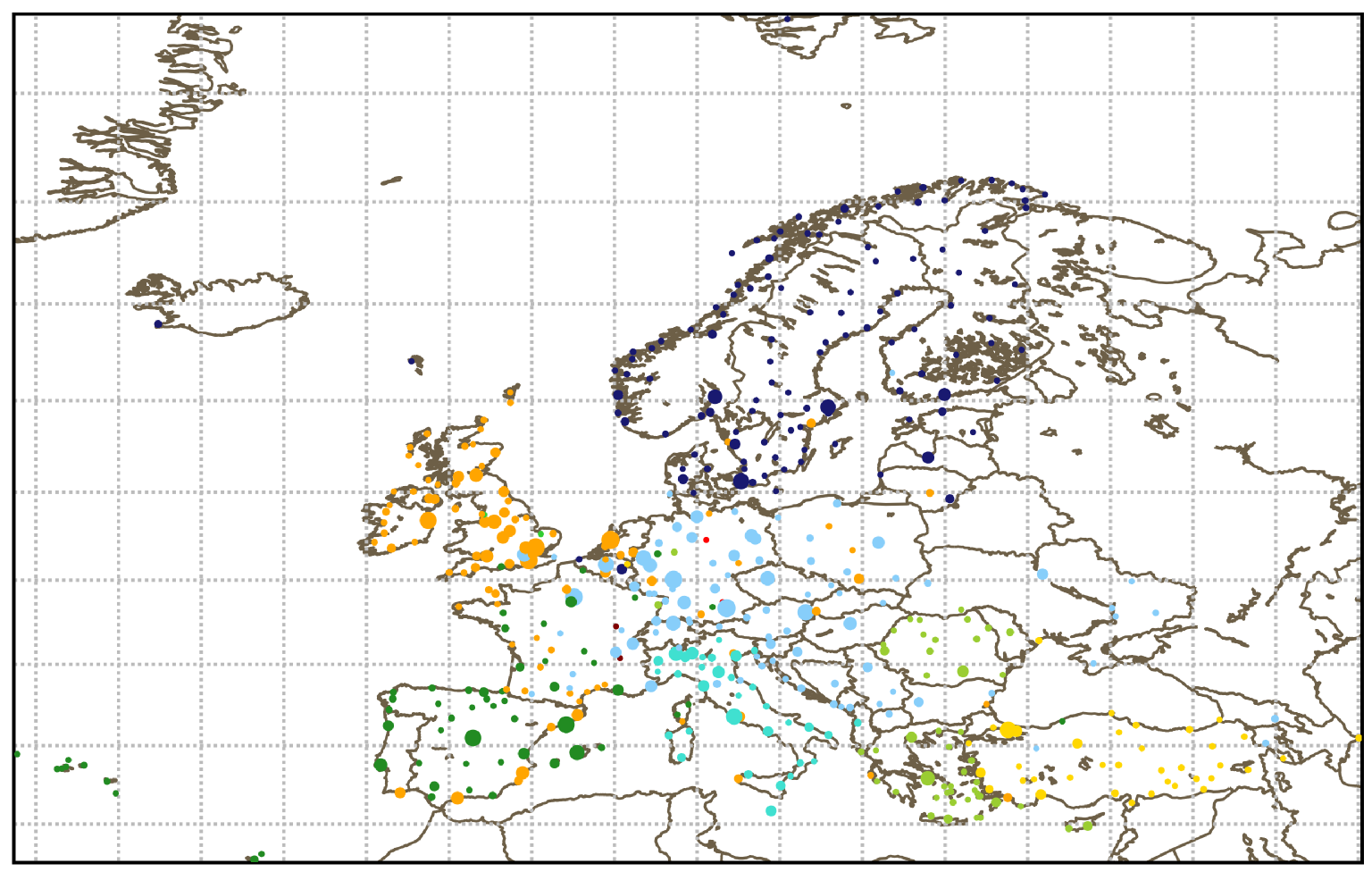}
\caption{European network of airports on May 6, 2010. Each circle is an airport, its radius proportional to its strength. Each community, detected with the modularity method, is represented by a different color. The links between nodes have been omitted for readability.}
\label{fig:communities_modularity}
\end{figure}

An example of a partition obtained with the modularity method is presented in Fig. \ref{fig:communities_modularity}. As one can see, the typical size of a community is supranational, roughly the same than a FAB. The communities are mainly geographical with the majority of nodes close to each other in a single community. Moreover, the borders of the communities seem to be more or less consistent with the national borders. Still, some nodes are far away from their communities.

The number of communities for each algorithm are presented in Table \ref{tab:nc_airports}. The modularity algorithm gives the biggest partitions, even bigger than the FABs. Instead, OSLOM and Infomap give between twice and thrice this number, but still less than the number of National Airspaces. In the second line of the table we show the minimum number of communities (induced or existing) which include 90\% of the nodes in the network. As for the sector network, we can see that  50\% of communities contains 10\% of nodes in the case of the NA partitioning.
\begin{table}[htbp]
\begin{tabular}{|c|c|c|c|c|c|}
\hline
Part. 		& Inf.			& Modularity	& OSLOM			& NA			& FABs \\
\hline
$N_c$		& 24.5.$\pm$3.9	& 9.4$\pm$1.2	&16.3$\pm$2.7	& 42.0$\pm$0. 	& 12.0$\pm$0.\\
\hline
$N_c^{0.9}$	& 12.0.$\pm$1.8	& 7.2$\pm$0.4	&10.2$\pm$1.0	& 21$\pm$0.4	& 9.1$\pm$0.3\\
\hline
\end{tabular}
\caption{Average number of communities in the network of airports and average minimum number of communities containing 90\% of the nodes. Computed over 28 days, with standard deviations taken as error bars.}
\label{tab:nc_airports}
\end{table}

Tables \ref{tab:airports_rand}, \ref{tab:airports_nmi}, and \ref{tab:airports_randA} show a summary of the procedure of comparison between unsupervised partitions and existing partitions. We see that all partitions share common features, since all values are far from 0. More specifically, the existing partition given by FABs seems quite well represented by the modularity method according to MI and RI. However, the match is obviously not perfect, and there could be two reasons for that: first,  geographical borders of communities are different from the FABs tiling; second,  communities are actually non geographical and some nodes of a given community are in the middle of another one, as shown in figure \ref{fig:communities_modularity}.
\begin{table}[htbp]
\centering
\begin{tabular}{|c||c|c|c|c|c|}
\hline
\bf{Rand Index}	& Infomap			& Modularity		& OSLOM 			& NA				& FAB\\
\hline	
\hline
Infomap			& 1.				&0.26$\pm$0.09		& 0.23 $\pm$0.01	&0.17$\pm$0.07		&0.22$\pm$0.05\\
\hline
Modularity		&0.26$\pm$0.09		& 1.				&0.54$\pm$0.06		&0.33$\pm$0.03		&0.41$\pm$0.03\\
\hline
OSLOM			& 0.23 $\pm$0.01	&0.54$\pm$0.06		&1.					&0.37$\pm$0.04		&0.36$\pm$0.04\\
\hline
NA				&0.17$\pm$0.07		&0.33$\pm$0.03		&0.37$\pm$0.04		&1.					& 0.64$\pm$0.009\\
\hline
FABs			&0.22$\pm$0.05		&0.41$\pm$0.03		&0.36$\pm$0.04		& 0.64$\pm$0.009	&1.\\
\hline
\end{tabular}
\caption{Comparison of the partitions using the Rand index. The errors bars are the standard deviations.}
\label{tab:airports_rand}
\end{table}

\begin{table}[htbp]
\centering
\begin{tabular}{|c||c|c|c|c|c|}
\hline
\bf{Mutual Information}	& Infomap			& Modularity		& OSLOM 			& NA				& FAB\\
\hline	
\hline
Infomap					& 1.				&0.54$\pm$0.04		& 0.50$\pm$0.07		&0.41$\pm$0.05		&0.47$\pm$0.04\\
\hline
Modularity				&0.54$\pm$0.04		& 1.				&0.60$\pm$0.04		&0.42$\pm$0.02		&0.53$\pm$0.02\\
\hline
OSLOM					& 0.50$\pm$0.07		&0.60$\pm$0.04		&1.					&0.45$\pm$0.04		&0.50$\pm$0.03\\
\hline
NA						&0.41$\pm$0.05		&0.42$\pm$0.02		&0.45$\pm$0.04		&1.					& 0.70$\pm$0.005\\
\hline
FABs					&0.47$\pm$0.04		&0.53$\pm$0.02		&0.50$\pm$0.03		& 0.70$\pm$0.005	&1.\\
\hline
\end{tabular}
\caption{Comparisons of the partitions using the mutual information. The errors bars are the standard deviations.}
\label{tab:airports_nmi}
\end{table}

The asymmetric Rand index of table \ref{tab:airports_randA} gives information on the inclusion relationships. We see that modularity partition, having the biggest communities, includes quite well all the partitions, except the Infomap partition. The latter seems to have boundaries not compatible with any other partition, since it is poorly included, even with the small sizes of its clusters. Note also that the OSLOM partition is much better included in the modularity one than in the NAs of FABs one. This means that OSLOM is a hierarchical refinements of modularity instead of the NAs or the FABs.

\begin{table}[htbp]
\centering
\begin{tabular}{|c||c|c|c|c|c|}
\hline
\bf{Asymmetric Rand Index}	& Infomap			& Modularity		& OSLOM 			& NA				& FAB\\
\hline	
\hline
Infomap						& 1.				&0.39$\pm$0.05		& 0.48$\pm$0.1		&0.58$\pm$0.07		&0.43$\pm$0.04\\
\hline
Modularity					&0.22$\pm$0.17		& 1.				&0.71$\pm$0.08		&0.52$\pm$0.06		&0.65$\pm$0.05\\
\hline
OSLOM						& 0.17$\pm$0.14		&0.44$\pm$0.06		&1.					&0.54$\pm$0.05		&0.34$\pm$0.04\\
\hline
NA							&0.1$\pm$0.07		&0.60$\pm$0.02		&0.28$\pm$0.04		&1.					& 0.48$\pm$0.01\\
\hline
FABs						&0.16$\pm$0.08		&0.71$\pm$0.02		&0.38$\pm$0.06		& 1.				&1.\\
\hline
\end{tabular}
\caption{Comparison of the partitions using the asymmetric Rand index. The errors bars are the standard deviations.}
\label{tab:airports_randA}
\end{table}

To have a more precise idea of the partitions and see if they stick to the FABs partition, we study the over-expression of nationalities in the communities of each partition. The results are shown in table \ref{tab:over_nat}. The communities detection highlights some of the main features of the future FAB partition. UK and Ireland, as well as Spain and Portugal, are always in the same community, which is a FAB. The northern countries are also together, although they are not exactly gathered the same way in all partitions. The central Europe is more problematic, Germany seems to play a central role there in reality, whereas it will be gathered with France in the same FAB. The role of France itself is also unclear in the community detection, probably because it acts as an important platform for Europe, thus not belonging clearly to a single geographical entity. Finally, while Italy and Greece are together in the same FAB, there is no partition where they are in the same community. In fact, probably due to its geographical location, Italy is always alone in its own community. On the contrary, it is striking to see that Greece can be in the same community than Romania and Turkey with Ukraine, whereas Greece and Turkey themselves are never in the same one. This might be due to the diplomatic issues between the two countries. Moreover, the FAB gathering Turkey, Azerbaijan and Georgia is visible in all partitions.

\begin{table}
\begin{tabular}{|c||c||c||c|}
	Modularity & OSLOM & FABs & Infomap\\
	\hline
	\hline
	\begin{tabular}{c|c|c|c}
Id & Size & Country & Frac.\\
\hline
\clt{0} & \clt{114} & \clt{Finland} & 0.94 \\ 
  		&  			& \clt{Denmark} & 1.0 \\ 
  		&  			& \clt{Norway} & 1.0 \\ 
  		&  			& \clt{Sweden} & 0.95 \\ 
\cl{1}  & \cl{97}  	& \cl{UK} & 0.90 \\ 
  		&  			& \cl{Ireland} & 1.0 \\ 
\clq{2} & \clq{90} 	& \clq{Germany C.} & 0.69 \\ 
\cld{3} & \cld{59} 	& \cld{Spain} & 0.85 \\ 
  		&  			& \cld{Portugal} & 0.92 \\ 
\clc{4} & \clc{49}	& \clc{Greece} & 0.97 \\ 
  		& 	 		& \clc{Romania} & 1.0 \\ 
\clqq{5} &\clqq{38} & \clqq{Turkey} & 0.92 \\ 
\cls{6} & \cls{33} & \cls{Italy} & 0.86 \\ 
\\
\\
\\
\\
\\
\\
\\
	\end{tabular}
	&
		\begin{tabular}{c|c|c|c}
Id & Size & Country & Frac.\\
\hline
\clq{0} & \clq{78}  & \clq{Germany C.} & 0.56\\ 
  		&  			& \clq{Austria} & 1.0 \\ 
\cl{1} 	& \cl{77} 	& \cl{UK} & 0.88 \\ 
  		&			& \cl{Ireland} & 1.0 \\ 
\clt{2} & \clt{73}  & \clt{Finland} & 0.94 \\ 
  		&  			& \clt{Sweden} & 0.90 \\ 
\cld{3} & \cld{45} 	& \cld{Spain} & 0.73 \\ 
  		&  			& \cld{Portugal} & 0.77 \\ 
\clqq{4}& \clqq{44} & \clqq{Turkey} & 0.92 \\ 
\clt{6} & \clt{41} 	& \clt{Norway} &  0.76  \\ 
  		& 	 		& \clt{Denmark} & 0.5\\ 
\cls{7} & \cls{33} 	& \cls{Italy} & 0.72 \\ 
\clh{8} & \clh{27} 	& \clh{Greece} & 0.79 \\ 
\clc{9} & \clc{18} 	& \clc{Romania} & 0.73 \\ 
\\
\\
\\
\\
\\
\\
	\end{tabular}
	&
	\begin{tabular}{c|c|c|c}
Id & Size & Country & Frac.\\
\hline
\clss{0}& \clss{90}	& \clss{Germany C.} & 1.0\\ 
  		& 			& \clss{France} & 1.0 \\ 
\cls{1}	& \cls{69}	& \cls{Greece} & 1.0 \\ 
  		& 			& \cls{Italy} & 1.0 \\ 
\clt{2} & \clt{66}	& \clt{Norway} & 1.0 \\ 
  		&			& \clt{Finland} & 1.0 \\ 
\cl{3}	& \cl{58}	& \cl{UK} & 1.0 \\ 
  		&  			& \cl{Ireland} & 1.0 \\ 
\clt{4} & \clt{47}	& \clt{Denmark} & 1.0 \\ 
  		& 			& \clt{Sweden} & 1.0 \\ 
\cld{5} & \cld{46}	& \cld{Spain} & 1.0 \\ 
  		& 			& \cld{Portugal} & 1.0 \\ 
\clqq{6}& \clqq{41} & \clqq{Turkey} & 1.0 \\ 
\clq{7} & \clq{22} 	& \clq{Croatia} & 1.0 \\ 
  		&			& \clq{Austria} & 1.0 \\ 
\clc{8} & \clc{18} 	& \clc{Romania} & 1.0 \\
9 & 13 & Poland & 1.0 \\ 
  & 13 & Lithuania & 1.0 \\ 
10 & 10 & Ukraine & 1.0 \\ 
11 & 6 & Serb. \& Mont. & 1.0 \\ 

	\end{tabular}
	&
	\begin{tabular}{c|c|c|c}
Id & Size & Country & Frac.\\
		\hline
\clss{0} 	& \clss{246}& \clss{Germany C.}	& 0.97\\
 			& 			& \clss{Spain} 	 	&1.0\\
 			& 			& \clss{France} 	&0.90\\
 			& 			& \clss{Italy} 	 	& 1.0\\
\clqq{1} 	& \clqq{38} & \clqq{Turkey} 	& 0.94\\
\clt{2} 	& \clt{30} 	& \clt{Sweden} 	 	& 0.72\\
\clh{3} 	& \clh{29} 	& \clh{Greece} 	 	& 0.93\\
\clt{4} 	& \clt{20} 	& \clt{Norway} 		& 0.44\\
\clc{5} 	& \clc{17}	& \clc{Romania} 	& 1.0\\
\clt{6} 	& \clt{14} 	& \clt{Norway} 	 	& 0.31\\
\clt{7} 	& \clt{12} 	& \clt{Finland} 	& 0.75\\
\clt{8} 	& \clt{10} 	& \clt{Norway} 	 	& 0.22\\
9 	& 9 	& Ukraine 	 	& 0.78\\
\clt{10} 	& \clt{9} 	& \clt{Denmark} 	& 0.88\\
\cl{11} 	& \cl{8} 	& \cl{UK} 			& 0.16\\
\cld{12} 	& \cld{8}	& \cld{Portugal} 	& 0.62\\
13 	& 7 	& Estonia 		& 1.0\\
14 	& 5 	& Serb. \& Mont. & 0.8\\
\\
\\
	\end{tabular}
\end{tabular}
\caption{Over-expressions of nationality in different partitions: for each partition, we display the ids of the communities, their sizes, the countries overexpressed within it and the fraction of the countries' airports included in the communities. The colours are guides to the eye, helping in identifying the main patterns. These partitions have been obtained for May, 6 2010.
}
\label{tab:over_nat}
\end{table}

The Infomap partition is also very different from the others. Indeed, it shows a massive community, gathering most countries in western Europe, including Ireland, UK, France, Benelux, Germany, Spain, Portugal, Italy, and  Switzerland, as well as countries in central Europe, including Austria, Croatia, Slovakia, Czech Republic, Slovenia, Hungary, and Poland. The community is in fact so big that some of these countries, even if they are totally included in it, are not overexpressed: the probability to find an airport of a given country within the community or in the whole Europe is almost the same. Only a few countries -- Greece, northern countries, Romania, Turkey -- are escaping this cluster, as well as a few western, small airports (less than 10). The massive community is present in several days of the AIRAC cycle, but not everyday.
\\

All these results seem to validate the general idea of the FABs, even if their actual boundaries could be different if based on the unsupervised community detection. However, the whole idea of inferring the FABs based on the communities of airports raise some issues. First, the airports \emph{are not} operated entities in the sense of the ATM, on the contrary of FABs. Of course, as a first approximation, one can consider that the traffic between airports give the main flow for the airspace. Second point, the airports can have long range interactions, and thus be a priori in the same community, while in two different countries which obviously cannot be in the same FAB. Hence, it is important to highlight these long range interactions and see how they are interfering with the rather ``compact'' communities we found in the network of airports.

\subsubsection{Extracting the role of distance in the airport network} \label{modgeo}

In  transportations systems like the airspace, nodes tend to be more connected to their geographical neighbors, just because they are close to each other. But of course there are other non local causes to the formation of communities. For instance, Ryanair has dedicated airports all over Europe, which are more likely to be connected. Hence, Beauvais in France (close to Paris) or Ciampino in Italy (close to Rome) are more likely to be connected to each other than to Fiumicino or Charles de Gaulle, respectively. 

\begin{figure}[htbp]
\centering
\includegraphics[width=0.6\textwidth]{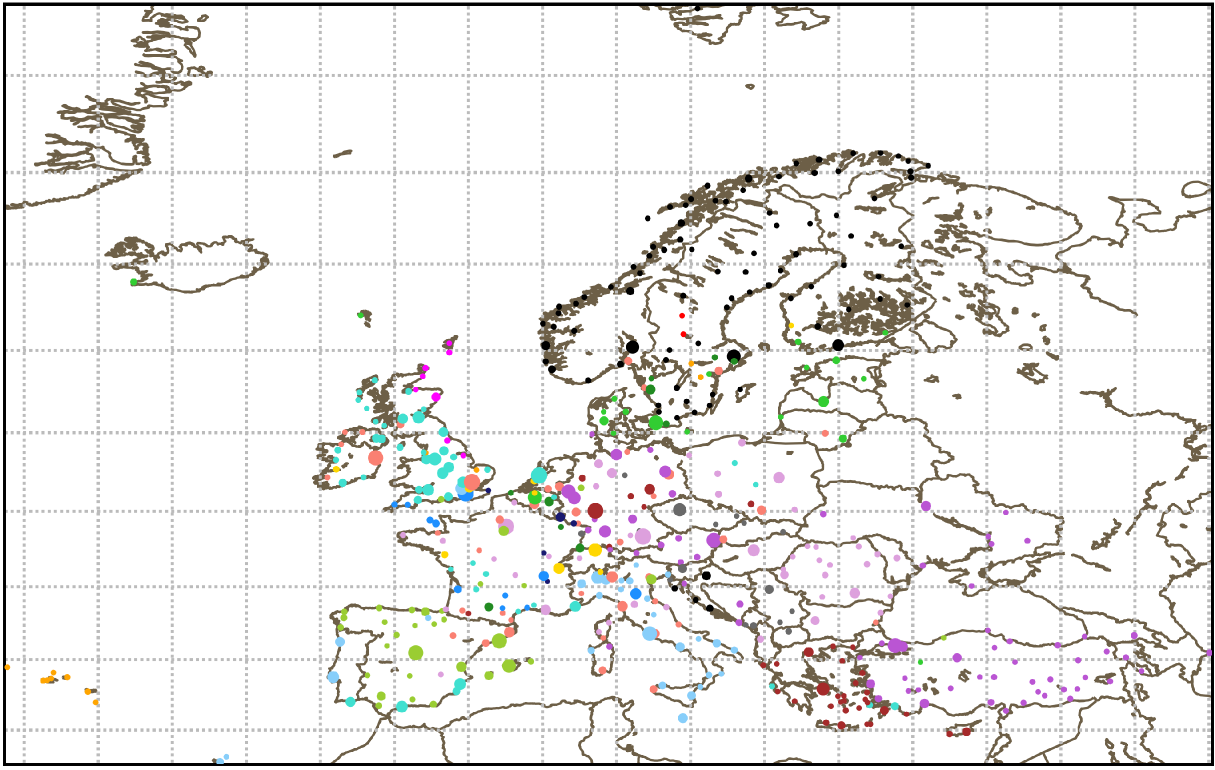}
\caption{Communities of the airport network obtained by maximizing the modularity with the spatial null model of Eq. (\ref{eq:spa}). The maximum is obtained by using a simulated annealing process.}
\label{fig:communities_spatial}
\end{figure}

A way of capturing this phenomenon and avoiding the distance bias has been described in section \ref{subsec:methods} and is based on the method described in Ref. \cite{spatial}. The idea is to maximize modularity by using a null hypothesis that takes into account the geographical distance. The result of this community detection is displayed in Fig. \ref{fig:communities_spatial}. As a point of comparison, one could consider  the partition of Fig. \ref{fig:communities_modularity}, which shows the modularity partition (with the usual null model) in the same day. The difference between the two figures is thus only in the null model (and the maximization algorithm).  

The picture with the geographical null model is very different. The communities are much less geographical constrained and long range interactions between airports are enhanced. For instance, the European airports operated mainly by Ryanair are now in the same community (displayed in salmon on the map). These interactions between these airports cannot be explained by geographical distance between them but only by the fact that they are operated by the same company. This community is not in the previous figure  \ref{fig:communities_modularity} and has been detected only by removing the distance bias.

In the same line, we can notice the specialization of airports surrounding big cities. Paris Orly for instance is in the same community than many Spanish airports, which is consistent with its actual role. Paris Charles De Gaulle is more linked to the central and eastern Europe and Beauvais, as noticed before, is operated only by Ryanair and is in its network. Around London each airport is also in a different community. Luton is more linked to other English airports, whereas Heathrow is linked to Italy and Portugal, Gatwick and the London City Airport to airports located in France and Switzerland mainly, and Stansted is in the Ryanair network. 

Let us emphasize also that geographical clusters did not disappear completely. Spain, United Kingdom, Italy, Greece and scandinavian countries have their own community, despite the spatial null model. This might be due to national characteristics (language, common firms, etc.) which are not directly concerned by distance. Moreover, the presence of bottlenecks, present for instance in the northern countries, can explain the clusters. Indeed, starting from a small airport, one has to go through Stockholm, Oslo or Helsinki to exit the country. Hence, regardless of the exact distance, all these airports are in the same community. 
\\



In conclusion the unsupervised detections of communities show two distinct characteristics of the network of airports. First, that it has a strong geographical basis, with airports more likely to be in the same community if they are close to each other. Moreover, the typical size of communities and also the geographical boundaries are consistent with the future partition of Europe. Secondly, the airports show some long distance interactions, which are not explained by geographical proximity, and cannot be strictly linked to the operational partitions. 

Another potential application to these detections may be to build direct dedicated communication tools between  airports that show a strong ``long-range interaction'' to smooth and optimize the coordination -- as already being addressed by some existing SESAR projects.

\section{Conclusions}\label{sec:conclusions}

We have investigated the structure of the ECAC airspace at different scales: the navigation point scale, the sector scale and the airport scale. We have used a network approach in order to highlight what are the links between the elements of the system, at each considered scale. We have then performed a community detection analysis in order to detect the groups of elements in the system that have homogeneous behaviour with respect to the actual air traffic conditions. Indeed, we have presented results relative to different unsupervised community detection algorithms that provide meaningful partitions of the airspace, starting from the mere knowledge of the actual air traffic flows. \\
At each scale the community detection algorithms provide useful insights on the system. In the case of the navpoint network we have obtained that the modularity partition gives big communities with sizes comparable to the NAs, while the OSLOM one gives a finer clusterization, close to the ACCs. The partitioning obtained by using the Infomap algorithm is close to reproduce the air traffic sectors. In the case of sector network the Infomap and OSLOM algorithms provide partitions which are quite close to the ACCs.  For the airport network, the considered algorithms provides a partition of the airspace close to the one of FABs. When the community detection algorithm is able to take into account the geographical constraints, as in section \ref{modgeo}, then some long distance interactions between airports emerge. These are not explained by geographical proximity, and cannot be strictly linked to general operational issues. Rather, they might reveal strategies operated by specific airlines, such as, for example, Ryanair. In fact, Ryanair has not a business model based on large hubs. Rather, flight plans are scheduled without explicitly considering the presence of connected flights.\\
Another interesting point is that, in the case of sector network, the modularity method, despite having a similar average size of communities, in not  close to the FABs partition. Thus, it is worth emphasizing that their boundaries are very different from the FABs ones, and this suggests a possible alternative design to the FABs, which preserves approximately their total number.\\
The different algorithms used are therefore able to capture different features of the airspace organization and, in some cases, they might provide alternative ways of designing the future European airspace. These new  partitions,  detected by mean of unsupervised methodologies, could be more optimized than the existing ones, because they follow a bottom-up approach based on actual traffic. The new ACCs would be indeed more densely connected inside and have less interface (links) with the exterior, which we believe it is an added value from an operational point of view. \\
All these results underline the difference between the bottom-up approach with an unsupervised partition based on the traffic and the top-down approach of the real construction of the existing partitions. \\
A detailed comparison between existing (e.g. natural) partitions and partitions obtained with network community detection could help to improve the design of airspace in different ways. For example, a local difference between the boundary of an existing sector and the corresponding community detected 
would suggest how the sector in the existing partition should be modified. \\
Another potential application is the use of these algorithms to highlight the boundaries (between sectors, ACCs, NAs, FABs) that require intensive coordination, as they may deserve dedicated coordination tools and procedures. The last application we mentioned is the establishment of direct communication links between closely-connected distant airports, as identified by the community detection applied to the airport network.
Potential areas of future studies may address the seasonal variations and suggest different airspace configurations to better cope with these demanding changes.\\
Our study is an example of how the community detection in networks provides an effective feedback on the appropriateness of the airspace design at the various scales considered, based on the sole knowledge of the actual air traffic data. It is important to emphasize that Europe (as well as the US) is moving toward a new scenario of air traffic management according to an ambitious and long term project termed SESAR \cite{sesar} aiming at changing the architecture of the European airspace based on a new set of paradigms \cite{sesar2007}. In this new scenario cross national control units will be defined and they will be based more significantly on traffic demand than on national constraints. In this framework our bottom up approach for the partitioning of the European airspace with community detection algorithms could be used to improve the design of this important transport infrastructure.\\
Finally, we also believe that this approach could be fruitfully adopted in other types of traffic networks \cite{invited}.

\section*{Acknowledgments}

We thank Marc Bourgois, Valentina Beato, and Alessandra Tedeschi for discussions and helpful comments. The work presented therein was co-financed by EUROCONTROL on behalf of the SESAR Joint Undertaking in the context of SESAR Work Package E project ELSA ``Empirically grounded agent based model for the future ATM scenario''. The paper reflects only the authors' views.  EUROCONTROL is not liable for any use that may be made of the information contained therein.

\appendix
\section{Existing partitions} \label{sec:appendix}

In this appendix we show the existing partitions present in the different networks, due to FABs, NAs, and ACCs. The FABs, or Functional AirBlocks, are described in table \ref{tab:compo_tabs}. We split the description in two parts. The first set contains the FABs which are already officially defined. The other set contains groups of countries (East Europe, Turkey, Serbia) that we defined according to geographical and cultural proximity in order to have a  perfect tessellation of ECAC airspace. Thus these are not actual FABs. 

\begin{table}[hbtp]
\centering
\begin{tabular}{c|c|c|}
Official FABs & Countries included\\
\hline
SW FAB 		& Spain, Portugal\\
FAB EC		& France, Switzerland, Monaco, Belgium, Germany, Netherlands, Luxembourg\\
Blue MED	& Italy, Cyprus, Greece, (Egypt, Tunisia), Albania, (Jordan)\\
FAB CE		& Czech Republic, Slovakia, Austria, Hungary, Croatia, Slovenia, Bosnia and Herzegovina\\
DANUBE		& Romania, Bulgaria\\
NEFAB		& Estonia, Finland, Latvia, Norway\\
UK-IR FAB	& United Kingdom, Ireland\\
Danish-Swedish FAB & Denmark, Sweden\\
BALTIC FAB	& Poland, Lithuania\\
\hline
\hline
Other FABs & Countries included\\
\hline
EAST Europe FAB	& Ukraine, Moldova, part of Russia\\
Turkey FAB & Armenia, Turkey, Azerbaijan, Georgia\\
Serbia FAB & Serbia and Montenegro, Macedonia\\
\end{tabular}
\caption{Composition of the Functional Airblocks (FABs).}
\label{tab:compo_tabs}
\end{table}

\begin{figure}[htbp]
\centering
\includegraphics[width=0.7\textwidth]{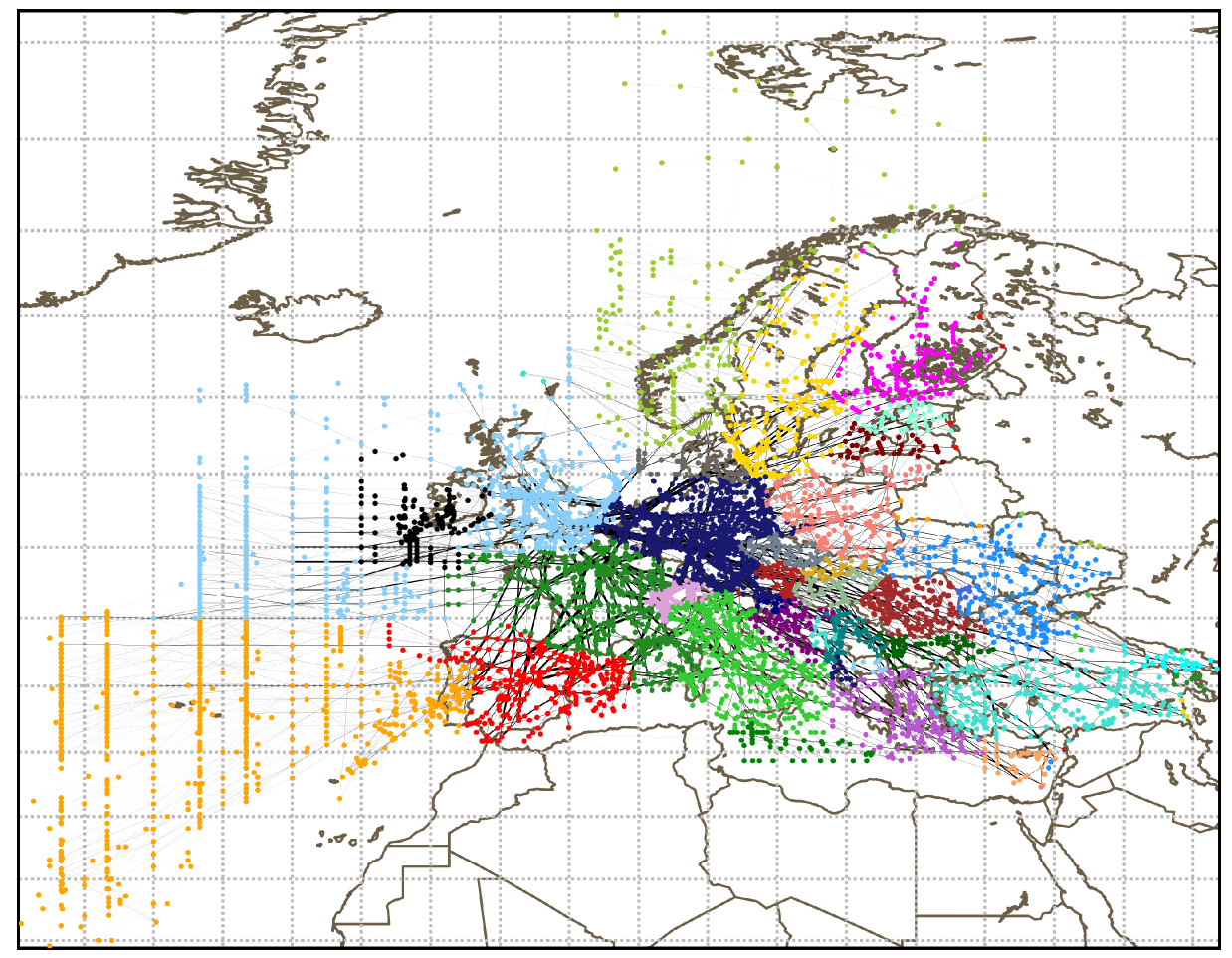}
\caption{Communities of the navpoint network based on the national airspaces.}
\label{navna}
\end{figure}

\begin{figure}[htbp]
\centering
\includegraphics[width=0.7\textwidth]{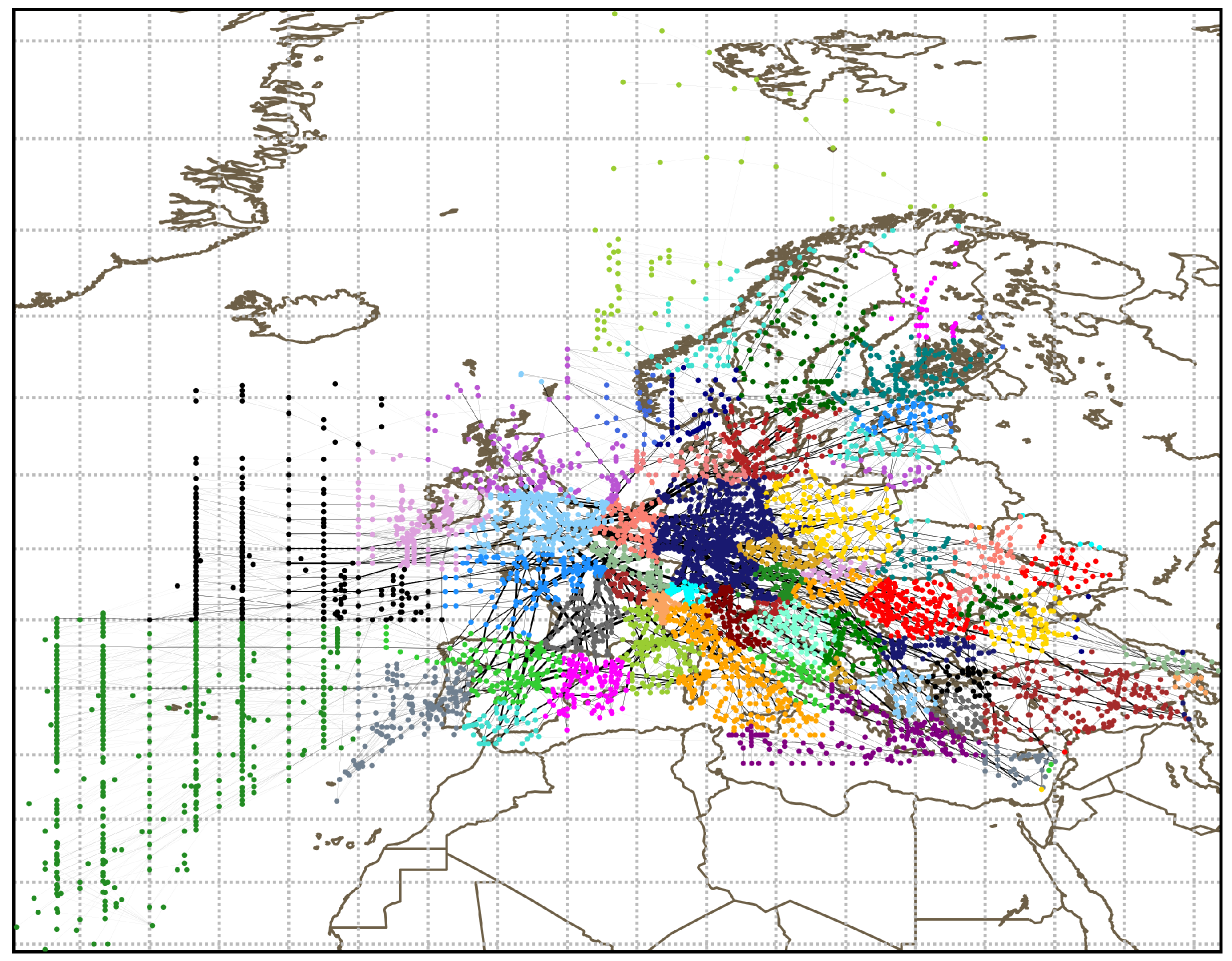}
\caption{Communities of the navpoint network  based on the control centres.}
\label{navacc}
\end{figure}

\begin{figure}[htbp]
\centering
\includegraphics[width=0.7\textwidth]{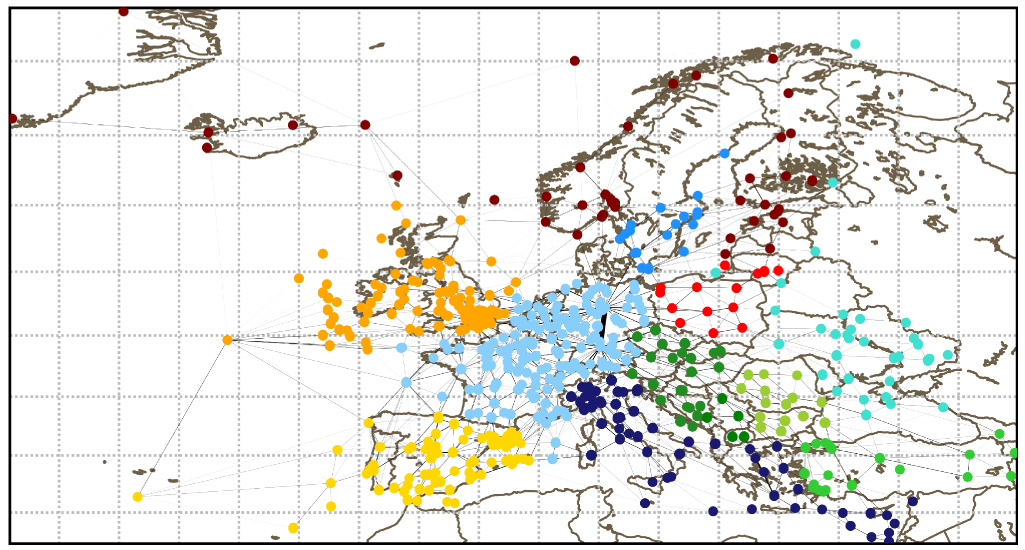}
\caption{Communities of the sector network based on the functional airblocks.}
\label{secfab}
\end{figure}

\begin{figure}[htbp]
\centering
\includegraphics[width=0.7\textwidth]{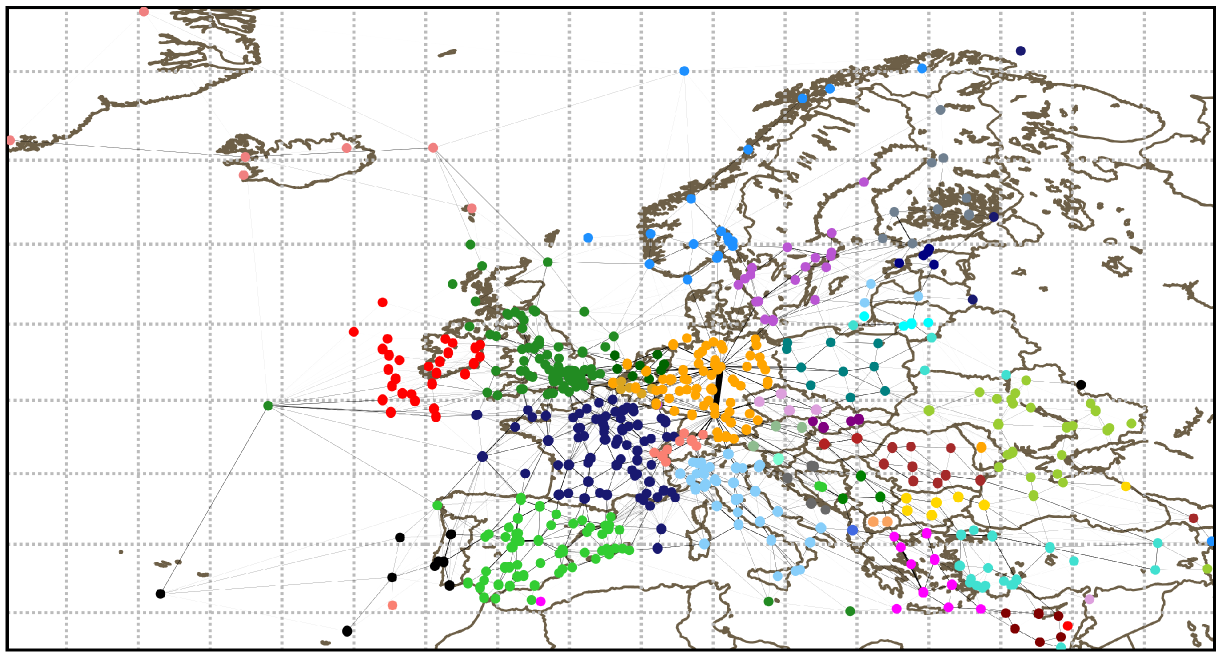}
\caption{Communities of the sector network based on the national airspaces.}
\label{secna}
\end{figure}

\begin{figure}[htbp]
\centering
\includegraphics[width=0.7\textwidth]{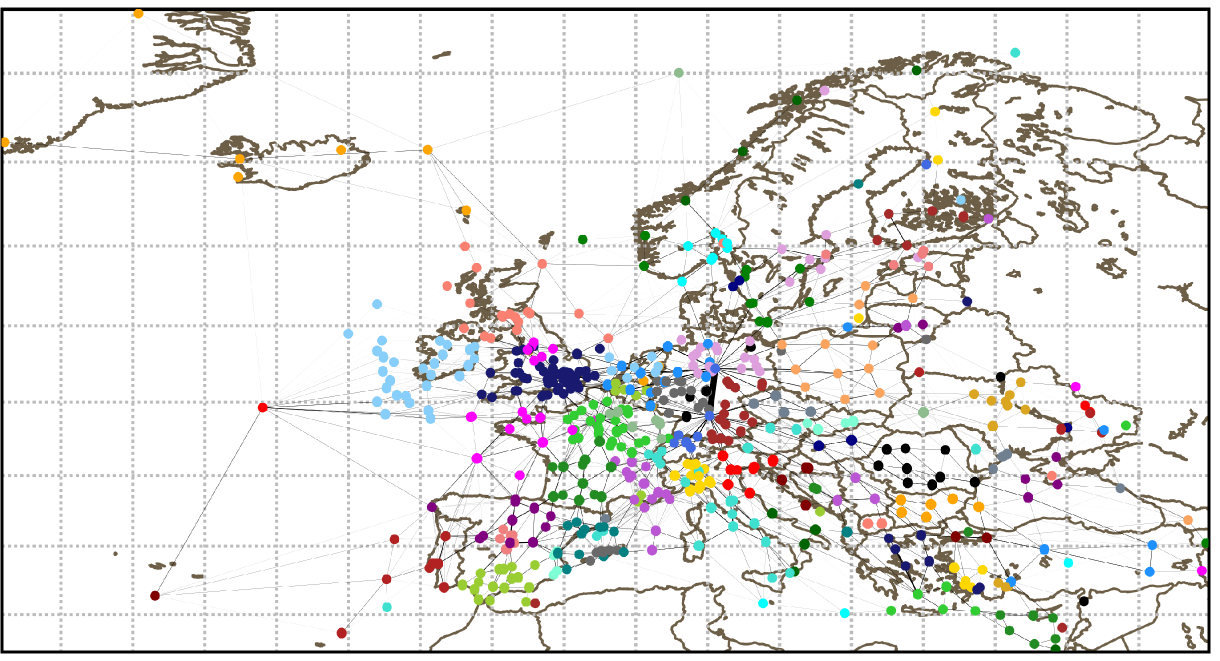}
\caption{Communities of the sector network based on the control centres.}
\label{secacc}
\end{figure}

\begin{figure}[htbp]
\centering
\includegraphics[width=0.7\textwidth]{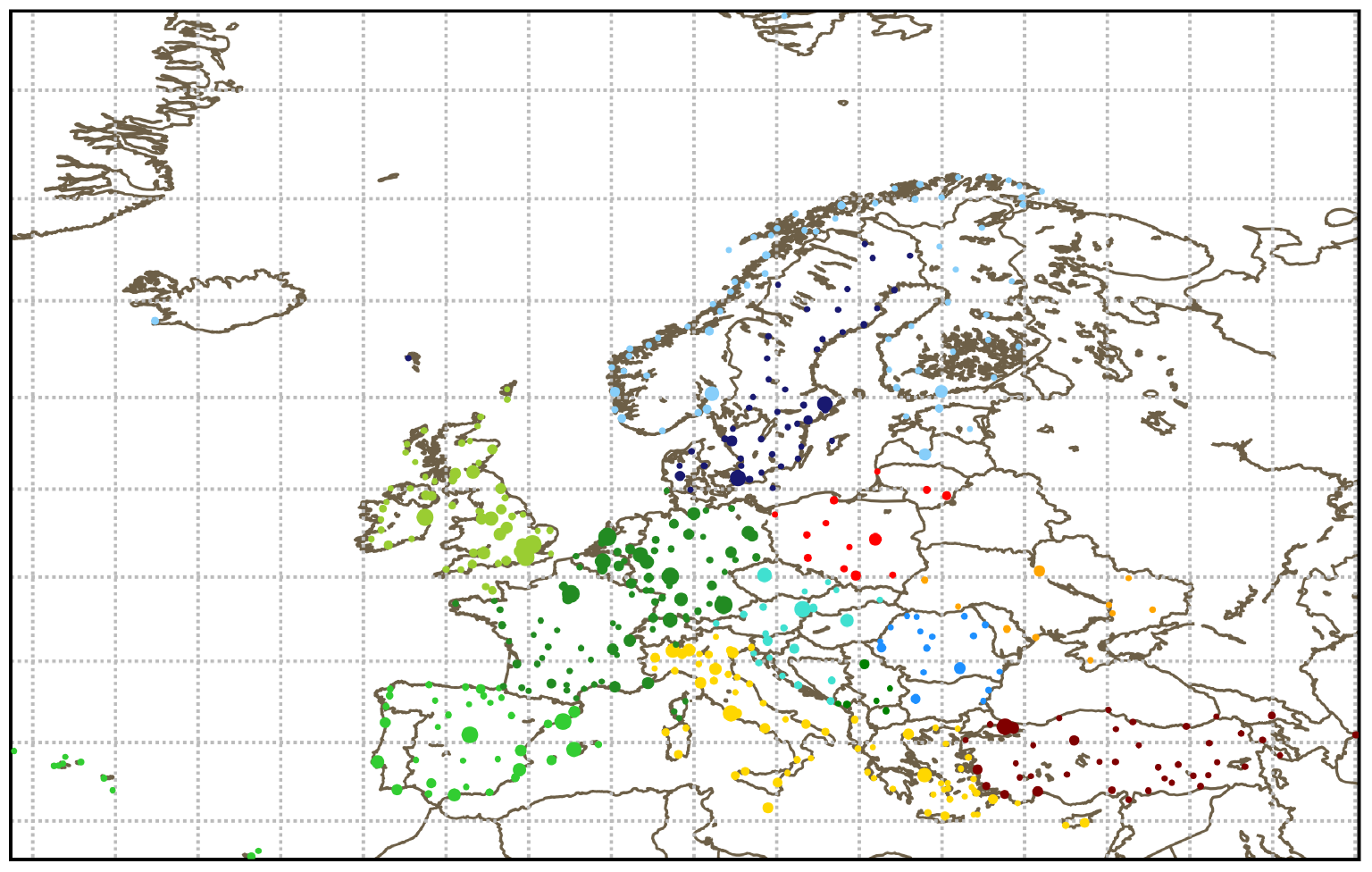}
\caption{Communities of the airport network based on the functional airblocks.}
\label{airfab}
\end{figure}

\begin{figure}[htbp]
\centering
\includegraphics[width=0.7\textwidth]{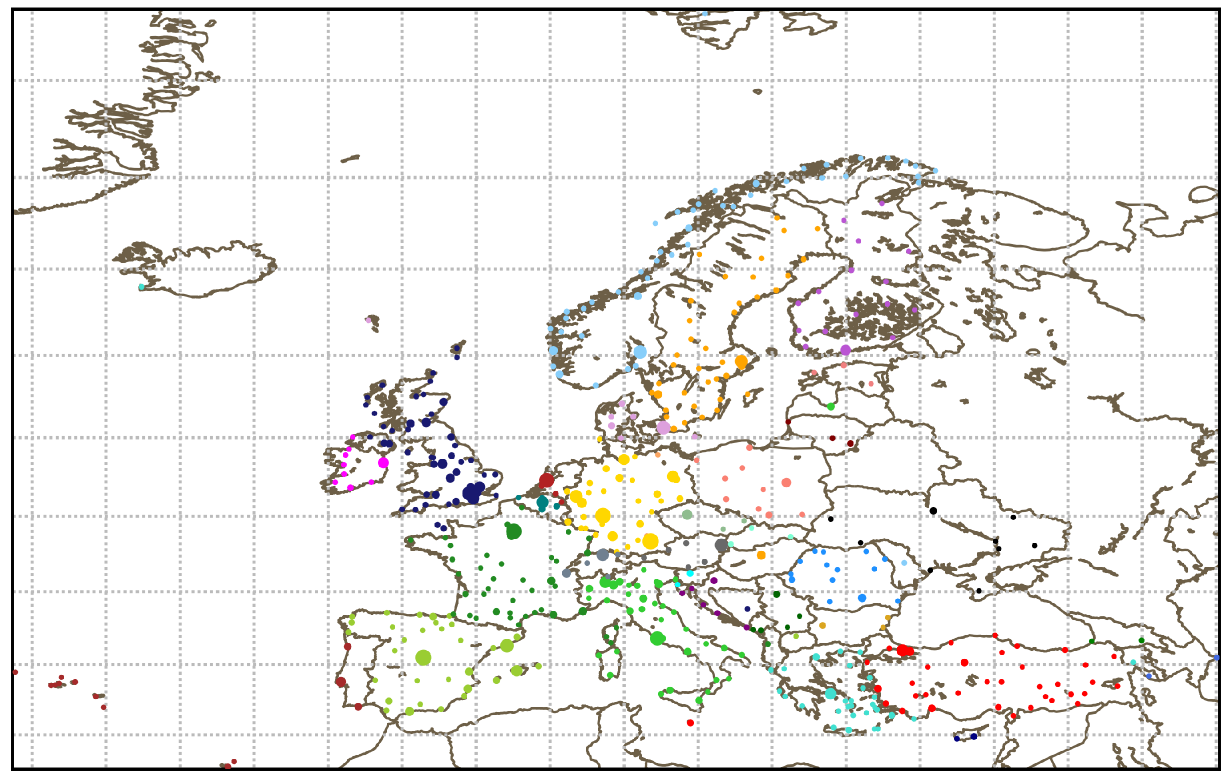}
\caption{Communities of the airport network based on the national airspaces.}
\label{airna}
\end{figure}

Here is a series of figures showing the different existing partitions of the systems.
\begin{itemize}
\item Figure \ref{navna} shows the communities of the navpoint network based on the national airspaces;
\item figure \ref{navacc} shows the communities of the navpoint network  based on the control centres;
\item figure \ref{secfab} shows the communities of the sector network based on the functional airblocks (FABs);
\item figure \ref{secna} shows the communities of the sector network based on the national airspaces;
\item figure \ref{secacc} shows the communities of the sector network based on the control centres;
\item figure \ref{airfab} shows the communities of the airport network based on the functional airblocks;
\item figure \ref{airna} shows the communities of the airport network based on the the national airspaces.
\end{itemize}


\end{document}